\documentclass[sigconf]{acmart}
\usepackage{amsmath,amsfonts,bm}

\def\eqref#1{equation~\ref{#1}}
\def\1{\bm{1}}

\DeclareMathAlphabet{\mathsfit}{\encodingdefault}{\sfdefault}{m}{sl}
\SetMathAlphabet{\mathsfit}{bold}{\encodingdefault}{\sfdefault}{bx}{n}

\newcommand{\sigmoid}{\sigma}

\usepackage[ruled,vlined]{algorithm2e}
\usepackage{import}
\usepackage{paralist}
\usepackage{subcaption}

\usepackage{caption}
\usepackage{multirow}
\usepackage{makecell}
\usepackage{threeparttable}
\usepackage{graphicx} 
\usepackage{enumitem}
\setlist[itemize]{leftmargin=*}

\AtBeginDocument{%
  \providecommand\BibTeX{{%
    \normalfont B\kern-0.5em{\scshape i\kern-0.25em b}\kern-0.8em\TeX}}}

\setcopyright{acmcopyright}
\copyrightyear{2018}
\acmYear{2018}
\acmDOI{10.1145/1122445.1122456}

\acmConference[XXX]{XXXX}{June 03--05, 2021}{XXXX}
\acmBooktitle{WXXXX,
  June 03--05, 2021, XXX, XX}
\acmPrice{15.00}
\acmISBN{978-1-4503-XXXX-X/18/06}

\graphicspath{./img}

\begin{document}

%%
%% The "title" command has an optional parameter,
%% allowing the author to define a "short title" to be used in page headers.

\title{Temporal aware Multi-Interest Graph Neural Network For Session-based Recommendation }

%%
%% The "author" command and its associated commands are used to define
%% the authors and their affiliations.
%% Of note is the shared affiliation of the first two authors, and the
%% "authornote" and "authornotemark" commands
%% used to denote shared contribution to the research.
% \author{Ben Trovato}
% \authornote{Both authors contributed equally to this research.}
% \email{trovato@corporation.com}
% \orcid{1234-5678-9012}
% \author{G.K.M. Tobin}
% \authornotemark[1]

\author{Qi Shen}
\authornote{Both authors contributed equally to this research.}
\affiliation{%
  \institution{Tongji University}
  \city{Shanghai}
  \country{China}}
\email{1653282@tongji.edu.cn}

\author{Shixuan Zhu}
\authornotemark[1]
\affiliation{%
  \institution{Tongji University}
  \city{Shanghai}
  \country{China}}
\email{2130768@tongji.edu.cn}

\author{Yitong Pang}
\affiliation{%
  \institution{Tongji University}
  \city{Shanghai}
  \country{China}}
\email{1930796@tongji.edu.cn}

\author{Yiming Zhang}
\affiliation{%
  \institution{Tongji University}
  \city{Shanghai}
  \country{China}}
\email{2030796@tongji.edu.cn}

\author{Zhihua Wei}
\authornote{Corresponding author.}
\affiliation{%
  \institution{Tongji University}
  \city{Shanghai}
  \country{China}}
\email{zhihua_wei@tongji.edu.cn}

% %

% \author{Qi Shen}
% \authornote{Both authors contributed equally to this research.}
% \affiliation{%
%   \institution{Tongji University}
%   \city{Shanghai}
%   \country{China}}
% \email{2130777@tongji.edu.cn}

% \author{Shixuan Zhu}
% \authornotemark[1]
% \affiliation{%
%   \institution{Tongji University}
%   \city{Shanghai}
%   \country{China}}
% \email{2130768@tongji.edu.cn}

% \author{Yitong Pang}
% \affiliation{%
%   \institution{Tongji University}
%   \city{Shanghai}
%   \country{China}}
% \email{1930796@tongji.edu.cn}

% \author{Yiming Zhang}
% \affiliation{%
%   \institution{Tongji University}
%   \city{Shanghai}
%   \country{China}}
% \email{2030796@tongji.edu.cn}

% \author{Zhihua Wei}
% \authornote{Corresponding author.}
% \affiliation{%
%   \institution{Tongji University}
%   \city{Shanghai}
%   \country{China}}
% \email{zhihua_wei@tongji.edu.cn}

% \author{Valerie B\'eranger}
% \affiliation{%
%   \institution{Inria Paris-Rocquencourt}
%   \city{Rocquencourt}
%   \country{France}
% }

\renewcommand{\shortauthors}{XXX, et al.}

%%
%% The abstract is a short summary of the work to be presented in the
%% article.

\begin{abstract}
%%background
Session-based recommendation (SBR) is a challenging task, which aims at recommending next items based on anonymous interaction sequences.
%disadvantage
Despite the superior performance of existing methods for SBR, there are still several limitations:
(i)  Almost all existing works concentrate on single interest extraction and fail to disentangle multiple interests of user, which easily results in suboptimal representations for SBR.
% Such uniform approach to model user interests easily results in suboptimal representations, failing to model diverse relationships and disentangle user intents in representations.
(ii) Furthermore, previous methods also ignore the multi-form temporal information, which is significant signal to obtain current intention for SBR.
%%solution
To address the limitations mentioned above, we propose a novel method, called \emph{Temporal aware Multi-Interest Graph Neural Network} (TMI-GNN) to disentangle multi-interest and yield refined intention representations with the injection of two level temporal information.
%%method in detail 
Specifically, by appending multiple interest nodes, we construct a multi-interest graph for current session, and adopt the GNNs to model the item-item relation to capture adjacent item transitions, item-interest relation to disentangle the multi-interests, and interest-item relation to refine the item representation.
Meanwhile, we incorporate item-level time interval signals to guide the item information propagation, and interest-level time distribution information to assist the scattering of interest information.
%%experiment  
Experiments on three benchmark datasets demonstrate that  TMI-GNN outperforms other state-of-the-art methods consistently.

%重要!: The abstract should briefly summarize the contents of the paper in
% 15--250 words.

%\and User Interaction.
\end{abstract}

%多加几篇引用

%%
%% The code below is generated by the tool at http://dl.acm.org/ccs.cfm.
%% Please copy and paste the code instead of the example below.
%%


%%
%% Keywords. The author(s) should pick words that accurately describe
%% the work being presented. Separate the keywords with commas.

%% A "teaser" image appears between the author and affiliation
%% information and the body of the document, and typically spans the
%% page.
% \begin{teaserfigure}
%   \includegraphics[width=\textwidth]{sampleteaser}
%   \caption{Seattle Mariners at Spring Training, 2010.}
%   \Description{Enjoying the baseball game from the third-base
%   seats. Ichiro Suzuki preparing to bat.}
%   \label{fig:teaser}
% \end{teaserfigure}

%%
%% This command processes the author and affiliation and title
%% information and builds the first part of the formatted document.
\maketitle

%贡献：
%1.通过引入时间模式来建模item之间的内部联系、用户当前兴趣的权重以及
%2.通过multi-head来建模兴趣簇？
%
%
%
%描述思路：

% 之前work的局限性：
%  1.session之前都旨在捕获用户的当前兴趣，而忽视了用户多view下的interest表征，没有做到disentangle
%  2.sequential 做disentangle的和session相关的是存在粒度差异的，长序列、长时间间隔的intention往往是独立的（第一天是电子产品、第二天是零食）、而session往往是一个基本intention下更细粒度的interest（比如不同的电子产品）
%  3. 之前基于item表征的intention解耦缺少其他辅助信息的帮助，而比如时间信息这种能更加准确的对兴趣簇进行抽取和反哺

%toy exmaple
%1. intention
%  1.1 每个物品存在多个interest，在session序列中interest会随着时间进行转移(短期的探索与尝试)
%  1.2 user的某些基本intention是保持不变的，故而整个session中也存在某些view下的interest重叠部分（贯穿全局的）
%
%2. time
%  2.1.对session中的每个item交互的时间分布不同，可能导致同样的交互序列，反映的是不同的用户意图；例如对一个长session而言，其中可能存在由较长时间间隔区分的多个明显兴趣区间，也可能为item间时间间隔都很短的、存在多个兴趣的单个区间。对前者需要对次要兴趣进行过滤，而对后者不能简单地将可能存在的多种兴趣进行杂糅，需要全面考虑。
%  2.2.探索和利用的patterns差异，探索可能的时间间隔比较长（需要更多的信息获取），user的next item可能也偏向于未见过的item或者其他item；而利用对应的时间间隔比较短（每个之间只是比较）但是item的重复频率更高，故而user的next item可能也更偏向于用户交互过的item
%
%
\vspace{-0.2cm}
\section{Introduction}
\vspace{-0.1cm}

% introduction of SBR
Recommender system has become the basis to relieve the information overload problem.
%for users. 
Most recommendation methods capture users' interest by modeling users' long-term preference for predicting their future interactions, e.g., collaborative filtering \cite{he2017neural,herlocker2004evaluating} and neural network based models \cite{DREAM,sedhain2015autorec}.
Differ from traditional recommendation tasks, in many practical scenarios, there is only a session available without access to user identification and historical interactions.
% For instance, we need to provide recommendation for users without login in e-commerce or media streaming sites.
This kind of task called session-based recommendation (SBR), aims to extract useful information as much as possible from limited data in current session.

Existing SBR methods mostly concentrate on modeling sequential information among items of current session by using Recurrent Neural Networks (RNNs) \cite{gru4rec,narm} or Graph Neural Networks (GNNs) \cite{srgnn,stargnn}.
However, these works simply regard a session as a short sequence with single intention, consider that the basic intention of user in a session usually remains the same, and try to capture user's current interest directly from the entire session. 
They overlook the fact that even in a relatively short-term session, the user’s finer granular interests can be multiple in various views, drift over time, or interweave with item overlaps.
Some studies \cite{DIN,MIND,DCF,DiverseTrend,CMI,Cap} in other recommendation areas have verified the effectiveness of modeling user's multi-interest, but in SBR, the multi-interest methods have not been fully explored.
% Different from in the long interaction sequence, the basic intention of user in a session usually remains the same, while finer granular interests may drift over time, or interweave with item overlaps.
As there is no interest disentangle mechanism in existing SBR methods, the mixture of major interests and minor interests may mislead the session representation learning so that it's hard to confirm users' true intentions and lacks of interpretability.
% For example, in \autoref{fig:toy-example}, the click sequence shows three different interests of user, i.e \textbf{Phone} product, \textbf{Apple} brand and \textbf{Laptop} product.
% Therefore, he may click items of the three categories during this period of time

% our example
\begin{figure}[t]
    \centering
    \begin{subfigure}{0.95\linewidth}
        \includegraphics[width=\textwidth]{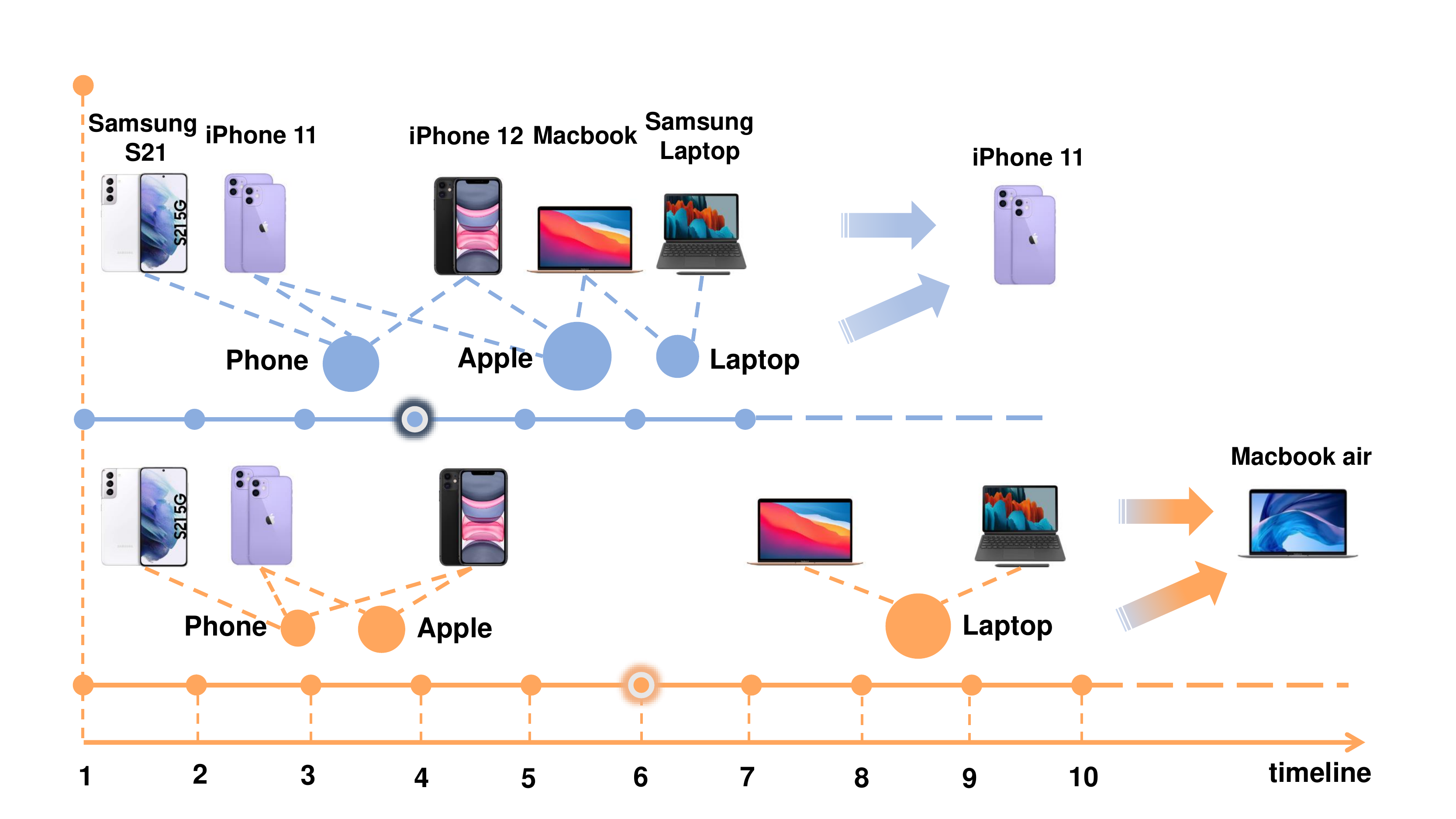}
    \end{subfigure}
    \caption{A toy example for two sessions with the same item sequence but different time intervals, where the circles connected to items denote the user's interests. Positions of the circles reflected the distribution of interest in the timeline, and the size indicates the importance of these interests.}
    %, where the color of transition arrow indicates the time interval width between items in session.
    %, warm colors represent relatively short intervals, and cold colors represent relatively long intervals.}
    \label{fig:toy-example}
    \vspace{-0.55cm}
\end{figure}

Moreover, the auxiliary temporal information is also ignored in session modeling.
Intuitively, two typical temporal information of interaction sequences can be recognized for multi-interest modeling:
% item-level time intervals information, interest-level semantic continuity information and session-level reverse sequence temporal information.
% (i) \emph{item-level time interval signals} is the intervals between item transitions, which reflect the relatedness between adjacent items.
% (ii) \emph{interest-level semantic continuity information} represents the interest distribution through virtual location and coverage in timeline,  which is helpful to model the relation between item and interest from the time perspective.
% % More importantly, interests transition can influence the distribution of such time intervals, so the transition time patterns reflect the importance of multiple interests.
% (iii) \emph{session-level temporal information} consists of intervals between last item and other items from last interaction to first one, which reveals the behavior pattern for exploration or exploitation.
% The tight time distribution of interaction patterns indicates the exploitation stage of user, while loose time distributions denote the exploration stage of user.
(i) \emph{item-level time interval signals} are the intervals between item transitions, which reflect the relatedness between adjacent items.
Generally, the tight time interval indicates the higher relevance of items, while loose time interval may mean the drifting of interest.
(ii) \emph{interest-level distribution information} represents the interest distribution through virtual location and coverage scope in timeline, which is helpful to model the relation between item and interest from the time perspective. For the chunked interest distribution in the session, the closer distance between items and interest factors in the timeline usually reveals the higher relatedness of them.
% More importantly, interests transition can influence the distribution of such time intervals, so the transition time patterns reflect the importance of multiple interests.

%importance of interests disentangle
%all the time intervals are short---- 
We illustrate this with an example in \autoref{fig:toy-example}, including two sessions with the same item sequence under different interaction time distributions. 
% basic intention of electronic products.
For the first session with short time intervals, as user's intention tends to be continuous over a short period of time, we can conclude that even item relatively far from current position may still represent user‘s strong interest. So user may mostly be interested in \textbf{Phone} product and \textbf{Apple} brand.
Under this circumstance, previous methods without introducing auxiliary time information will regard it as a session with evenly and moderate length time intervals.
Therefore, a deviation from the major interests may be caused by the latest items, which are laptops.
% Moreover, once the distribution of time interval changes, it's a different story. 
While for the second session with a relatively long time interval between \emph{iPhone} and \emph{Macbook}, we argue that the interest drift may occurred during the interval.
Thus the interests distributed in the front portion of timeline such as \textbf{Phone}, ought to be a minor factor for recommendation, while the interest \textbf{Laptop} in the latter portion should be taken into special consideration.
But for SBR methods ignoring time information, the multiple interests in two chunks of the session cannot be effectively disentangled and rearranged.
% Affected by previous item \emph{Phone11},
The minor interest \textbf{Phone} is likely to be over weighted for user's current intention, while the crucial interest \textbf{Laptop} may be diluted.
As discussed above, even sessions with exactly the same items and orders, but for diverse time distributions, may have diverse interests intensity, leading to completely different next items.
In this case, the ignorance of time information makes existing methods insufficient to distill effective intention signals from the active session.

% our proposed 
To address these problems, we propose \emph{Temporal aware Multi-Interest Graph Neural Network} (TMI-GNN), a novel method to distill the disentangled interest and inject the temporal information for better inferring the user intentions of the current session.
To be more specific, we firstly construct a multi-interest graph for current session by appending multiple interest nodes into original item-item graphs, which builds the potential relations between items and different interest factors.
Then, to generate the item and interest representations, we synchronously apply item information propagation with item-level time interval signals, interest extraction in a soft clustering way, and interest attaching with the interest-level distribution information.
Finally, we integrate item embeddings and interest embedding under the guide of temporal distance, to represent user’s preference for next item prediction.
% The effectiveness of our proposed SBR model is verified through extensive experiments on three real-world datasets.

Our main contributions of this work are summarized below:
\begin{itemize}[leftmargin=*]
    \item We propose to construct a multi-interest graph with item and interest nodes for representing multi-interest session effectively, involving the explicit item transitions and potential connections between different items and interests.
    \item For the constructed graph, we develop a novel \textbf{TMI-GNN} model for SBR to capture adjacent item transitions, distill users’ current multi-interests from noisy interaction sequences and feedback related interest information to enhance session representations. Moreover, it explicitly injects item-level and interest-level temporal information into the above process to refine current intention representation.
    \item Extensive experiments on three datasets demonstrate that our model is superior compared with state-of-the-art models.
\end{itemize}

%如果要缩为12页，这部分可以大大删减
%multi-categories部分可以和session部分合并

\vspace{-0.1cm}
\section{Related Works}
\vspace{-0.05cm}
\textbf{Session-based Recommendation.}
Following the development of deep learning, many neural network based  approaches have been proposed for SBR.
% \emph{RNN-based SBR.}
Due to well sequence modeling capability of RNNs,  RNN-based methods have been widely used for SBR \cite{gru4rec,narm,liu2018stamp,ren2019repeatnet,song2019islf}. 
For instance, GRU4Rec \cite{gru4rec} was firstly proposed to utilize GRU layer to capture information in interaction sequences. 
Based on GRU4Rec, NARM \cite{narm} added an attention mechanism after RNN, which refers to last interaction item, in order to capture the global and local preference representation of user in current session.
But in an ongoing session, interaction patterns are always more complex than simple sequential signals, which cannot be effectively captured by RNN-based models.
% \emph{GNN-based SBR.}
More recently, motivated by the superior performance of GNNs in extracting complex relationships between objects, quite a few recommendation methods relying on GNNs were proposed to extract the item transition patterns for SBR \cite{srgnn,xu2019graph,fgnn,stargnn,lessr,gce-gnn,pang2021session,DATMDI}.
For example, SRGNN \cite{srgnn} converted the interaction sequence into a directed graph and employed the gated GNN (GGNN) on the graph to learn item embedding.
% However, the GNN-based methods above propagate information between adjacent items, failing to consider long-distance item relations.
LESSR \cite{lessr} formed better graph structure from the session by proposing a lossless encoding scheme, and proposed a SGAT layer to model the long-range dependency, which propagates information along shortcut graph.
StarGNN \cite{stargnn} put forward a  star graph neural network to model the complex transitions between items with additional node to connect the  long-range item relations.
DATMDI \cite{DATMDI} learned the cross-session and local session representations via the GNN and GRU and combined them by a soft-attention mechanism.
% Both of the two methods solve the ineffective long-range dependency capturing problem.
% 

\vspace{-0.02cm}
\noindent \textbf{Temporal information in Recommendation.}
Meanwhile, temporal information is also a key feature in many recommendation scenarios, such as e-commerce and  video recommendation.
There are a few works to utilize the temporal information as contextual feature for recommendation \cite{vassoy2019time,li2020time,Wu2017Modeling,gu2020hierarchical}.
% \cite{Wu2017Modeling} expanded latent factor model, introduced the temporal information as states, and characterized the dependency and transition between users’ current latent vector via social networks.
However, most existing methods simply reduce the temporal information to order relationship, subsequently use RNN-based model to capture the sequential signal.
For instance, \cite{DREAM} used RNN to learn a dynamic representation of a user which reveal user’s dynamic interests at different time in next basket recommendation.
\cite{timematters} discovered absolute time patterns and relative time patterns based on insightful data analysis to model users’ temporal behaviors for recommender systems.
\cite{zhou2018micro} proposed a framework called RIB, which takes dwell time into account.By using the time a user spends on one item as a part of so called micro behavior, it can model more practical user intent.
% Similarly, \cite{gu2020hierarchical} used a behavior LSTM to model the dwell time and interval time during hierarchical user profiling construction process to enhance session modeling.

\begin{figure*}
\vspace{-0.2cm}
    \centering
    % \includegraphics[width=\linewidth]{img/overview.pdf}
    % \scriptsize
    % \def\svgwidth{0.7\linewidth}
    % \import{img/}{framework.png}
    \includegraphics[width=0.80\textwidth]{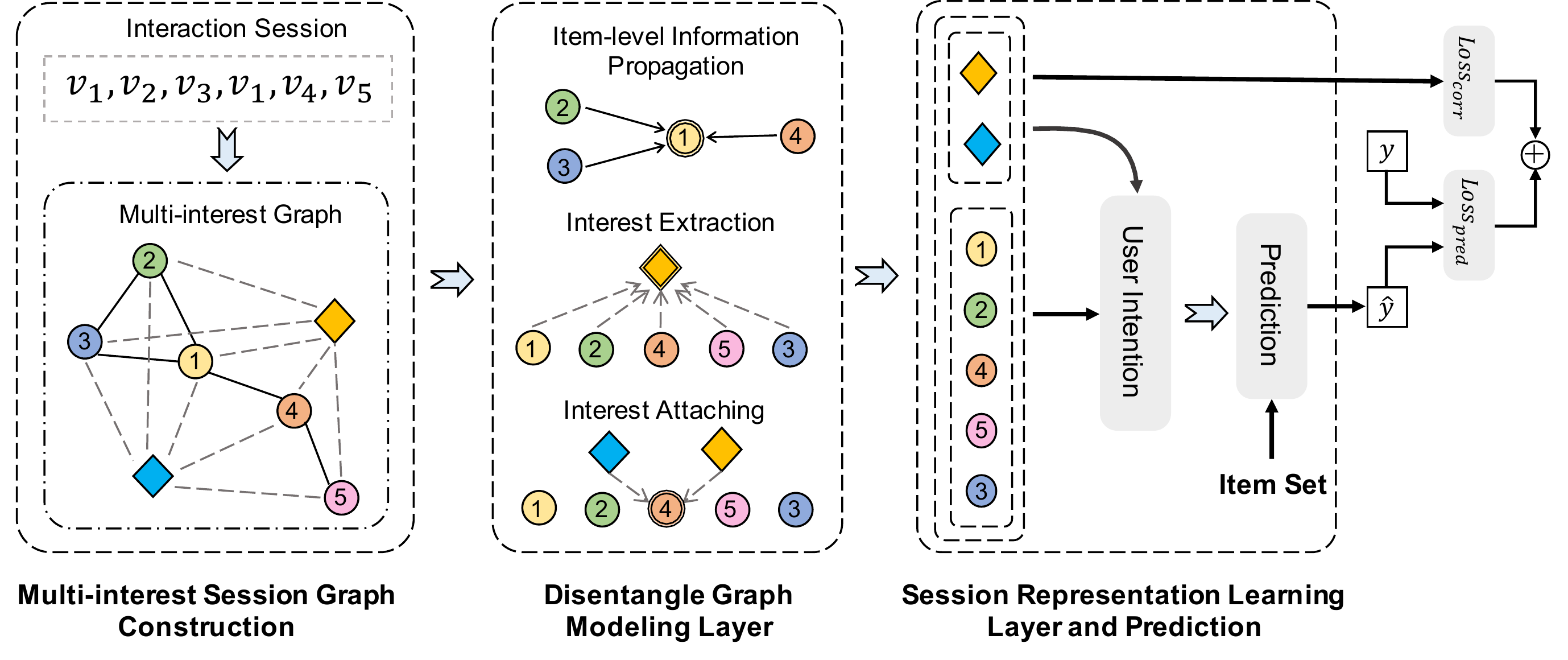}
    \caption{The overview of TMI-GNN. 
    The multi-interest graph contains three types of relations: \emph{item to item}, \emph{item to interest} and \emph{interest to item}.The Disentangled Graph Modeling layer learns the interest and item embeddings under the guide of rich temporal information. Finally, we aggregate the interest and item representations to generate the final session embedding. Moreover, an additional interest independent loss is considered to encourage the diversity of interests. }
    \label{fig:framework}
    \vspace{-0.2cm}
\end{figure*}

As outlined above, previous works on SBR have some limitations. 
% due to a lack in good means to bring temporal information into the session graph structure,
First, temporal information is rarely or crudely exploited in these works.
What's more, none of these methods explicitly disentangle the multiple interests of user in a session.
These limitations may lead to suboptimal performance.
%模型

\vspace{-0.1cm}
\section{Preliminary}\label{sec:define}
\vspace{-0.05cm}
In this work, we aim to explore the effectiveness of abundant temporal information and multiple disentangled interest for SBR.
Given the entire item set $\mathcal{V}$, we first define a timestamp-augmented item session sequence as $s=\{(v_1,t_1),(v_2,t_2),...,(v_{L},t_{L})\}$, where $(v_i,t_i)$ represents the user interacted item $v_i \in \mathcal{V}$  at time $t_i$ and $t_i < t_{i+1}$, and $L$ is the session length.
Given session $s$, the goal of SBR is to predict a probability score for any item $v \in \mathcal{V} $ such that an item with higher score is more likely to be interacted next.

\vspace{-0.15cm}
\section{Methodology}\label{sec:model}
\vspace{-0.05cm}
\subsection{Overview}
\vspace{-0.05cm}

In this section, we detail the design of our model.
As shown in \autoref{fig:framework}, it contains four main components: Multi-interest Session Graph Construction, Disentangle Graph Modeling Layer, Session Representation Learning Layer and Prediction Layer.
By constructing item-transition sequences as multi-interest graphs with additional interest nodes, we explicitly build the potential connection between latent user preference and explicit items sequence.
Furthermore, at the Disentangle Graph Modeling Layer, we extract item transition information, distill multi-interest representations and feedback disentangled interests to related item embeddings, respectively.
With the guidance of item-level and interest-level temporal information, the refined attention is better estimated for disentangling user's multi-interest.
Then, Session Representation Learning Layer adaptively generates the final intention representations with the injection of last-item based time interval information.
Finally, the predictor estimates the probability of candidate items based on the each disentangled session representations.

\vspace{-0.15cm}
\subsection{Multi-interest Session Graph Construction}\label{sec:graph_cons}
\vspace{-0.05cm}
% As mentioned above, the adjacent item transitions graph restricts the capabilities of models to capture long-range dependencies.
As mentioned above, the extraction of multi-interest is meaningful for obtaining user intention representation from the interaction sequence.
% Hence, we consider adding a node with connections to all items to bridge the gap of long-range relations of items for current session. 
% However, the incorporation of single node into session graph may restrict the capabilities of interest representation and confuses the final item embeddings after multiple GNN layers. 
% Therefore, we consider not only the item nodes based on adjacent item transitions in current session, but the multiple interest nodes representing the interest factors with full connection to item nodes, to construct the undirected multi-interest session graph
Therefore, we propose multi-interest session graph for effectively organizing the interactions in the session and modeling user's multi-interest.

For given session $s$, we construct the multi-interest session graph $\mathcal{G}_s=\left(\mathcal{V}_s,\mathcal{E}_s\right)$ ,where $\mathcal{V}_s=\left(\{v_1,v_2,v_3,\dots,v_N\}, \{u_1,u_2,\dots,u_H\}\right)$ indicates the node set of the constructed graph which contains $N$ item nodes and $H$ interest nodes.
Besides the basic transition between items in the session, we additionally introduce the \textbf{interest nodes} to represent each independent interest, which can be explained as different distributions of items' contribution to user's intention.
For $j$-th interest node, it fully connects to all items in the session with edges $\{(u_j,v_i) | 1\leq i \leq N\}$.
For each session, each item node $v_i$ has corresponding item-interest edges $\{(v_i,u_j) | 1\leq j \leq H\}$ to each interest node, and item-item transition edge to contextual item nodes.
Through the full connection between the explicit item and interest node, the soft assignment of each item to corresponding interest can be estimated by subsequent GNN as edge attribute.

\vspace{-0.3cm}
\subsubsection{Temporal Information.}
Furthermore, we attach auxiliary multi-form temporal information to the multi-interest graph for distilling more precise interest representations.
For the original timestamp sequence $(t_1,t_2,\dots,t_{L})$ of session $s$, item-level transition interval $t_{ij}$ is attached to the edge of item $i$ and $j$:
% and the timestamp $t_{-i}$ is attached to the interaction position $L-i+1$:
\begin{equation}
\setlength{\abovedisplayskip}{3pt}
\setlength{\belowdisplayskip}{3pt}
\begin{aligned}
     t_{i,j}=\frac{\left|t_{i}-t_{j}\right|}{t_{bucket}} \,,
% \;;\;
    %  t_{i}=\frac{t_{i}-t_{1}}{t_{bucket}}
\end{aligned}
\end{equation}
where $t_{bucket}$ denotes the pre-defined length  of time bucket.
Moreover, the relative time-step $t_{i,1}$ is attached to the interaction position $i$, compared to the start time at the first position $1$.

For the constructed multi-interest session graph, there are three types of relation, i.e., item-item, item-interest and interest-item relation, and we represent them by superscript $v\xrightarrow[]{}v$, $v\xrightarrow[]{}u$ and $u\xrightarrow[]{}v$ respectively.

\vspace{-0.15cm}
\subsection{Disentangle Graph Modeling Layer}
\vspace{-0.05cm}
% 1. 接下来，我们介绍在multi-interest graph进行多兴趣建模，即兴趣解耦操作。
%2. 我们基于message传播-聚合的范式从图中更新物品/兴趣node的表征 % 3. 通过interest node之间item node边的权重分配，完成兴趣提取和解耦
% Next, we present how to propagate features on the heterogeneous multi-interest graph to encode item-transition information and disentangle latent multi-interest with the auxiliary multi-form temporal information.
Next, we present how to obtain item representations and disentangle interest representations on the constructed heterogeneous multi-interest graph.
Based on the message propagation of GNNs, the item and interest node embeddings are updated based on the previous results with neighbor information.
The disentanglement of multi-interest can be consider as the process of iteratively refining interest node embedding, and the explainable assignments for each interest factor can be estimated as the weight of item-interest edges based on the full connection of each independent interest node and the item nodes, i.e., each interest factor is generated by item information pooling with different assignment scores.
For the adjacent item transition, previous GNN-based model, like SRGNN \cite{srgnn}, has achieved superior performance for SBR.
Therefore, for each GNN layer, we firstly aggregate the neighbor messages in each relation respectively.
Then we gather the semantic information to update the item and interest representations.

Let $\mathbf{u}_{i}^{(k)},\mathbf{v}_{j}^{(k)}$ denote the embedding of interest $u_i$ and item $v_j$ after $k$ layers GNN propagation.
The item IDs are embedded into $d$-dimensional space and are used as initial node features in our model, $\mathbf{v}_{j}^{(0)}\in \mathbb{R}^d$. For the multiple temporal information, we embed them by a learnable temporal matrix $\mathbf{T} = [\mathbf{T}_{0}, \mathbf{T}_1, \mathbf{T}_2, \dots, \mathbf{T}_m]$ for the rounded time value $0\leq t \leq m$, where $m$ is the max time-step intervals. 

Moreover, for each interest factor as related interest node, we adopt average operation on the item nodes to initialize the interest representation, i.e., $\mathbf{u}_{i}^{(0)}=\frac{1}{L}\sum_{j=1}^{L}\mathbf{v}_{j}^{(0)}$.
% \begin{equation}
% \setlength{\abovedisplayskip}{3pt}
% \setlength{\belowdisplayskip}{3pt}
% \begin{aligned}
%     \mathbf{u}_{i}^{(0)}=\frac{1}{L}\sum_{j=1}^{L}\mathbf{v}_{j}^{(0)}\,,
% \end{aligned}
% \end{equation}
% where $L$ is the sequence length of activate item session.
For the interest-side temporal information, we utilize the the center timestamp $t_{cent,i}$ to indicate the location of $i$-th interest factor on the timeline, and the temporal compactness value $t_{comp,i}$ to indicate the coverage of interest factors in the item sequence.
Here we initialize these two characteristics of each interest factor with average pooing, i.e., $t_{cent,i}^{(0)}=\frac{1}{L}\sum_{j=1}^{L} t_{j,1}$, and $t_{comp,i}^{(0)}=\frac{1}{L}\sum_{j=1}^{L} |t_{j,1}-t_{cent,i}^{(k)}|$.
%to measure the midpoint and deviation of interest distribution.
% For the interest-level temporal information, center timestamp and temporal compactness value are generated based on the learnt attention coefficients and item timestamp, which will be detailed in \autoref{sec:i2v}.
% The center timestamp indicates the  location of interest factors on the timeline, and the temporal compactness value indicates the coverage of interest factors to item sequence in temporal view.
% \begin{equation}
% \begin{aligned}
%     t_{cent,i}^{(0)}=\frac{1}{L}\sum_{j=1}^{L} t_{-j}\,.
% \end{aligned}
% \end{equation}
% \begin{equation}
% \begin{aligned}
%     t_{comp,i}^{(0)}=\frac{1}{L}\sum_{j=1}^{L} |t_{-j}-t_{center}^{(k)}|\,.
% \end{aligned}
% \end{equation}

Then, we will detail the modeling for item-item, item-interest and interest-item relations, respectively.
% To learn a cross-semantic representation according to the multi-interest graph, we update the interest node and item node at each GNN layer as follows.
\vspace{-0.15cm}
\subsubsection{Item-level Information Propagation Layer}\label{sec:i2i}
For the item-item relation, we adopt SRGNN \cite{srgnn} with detail time interval information to propagate adjacent node information. 
It assembles neighbor node information with the normalized coefficient  $e_{ij}^{v\xrightarrow[]{}v}$ under the guide of time interval signal $t_{ij}$: 
\begin{equation}
\setlength{\abovedisplayskip}{3pt}
\setlength{\belowdisplayskip}{3pt}
\begin{aligned}
    e_{ij}^{v\xrightarrow[]{}v}=\underset{j\in \mathcal{N}_{v\xrightarrow[]{}v}(i)}{\text{softmax}}\left(\text{MLP}\left(\mathbf{T}_{t_{ij}}\right)\right) \,,
\end{aligned}\label{eqn:grad}
\end{equation}
\begin{equation}
\setlength{\abovedisplayskip}{3pt}
\setlength{\belowdisplayskip}{3pt}
\begin{aligned}
    \mathbf{m}_{i}^{(k)}=\sum_{j \in \mathcal{N}_{v\xrightarrow[]{}v}(i)}  e_{ij}^{v\xrightarrow[]{}v} \mathbf{v}_{j}^{(k)} \,,
\end{aligned}\label{eqn:grad}
\end{equation}
% \begin{equation}
% \begin{aligned}
%     \mathbf{m}_{i}=\frac{1}{D\left(i\right)}\sum_{j \in \mathcal{N}_{v\xrightarrow[]{}v}(i)}\mathbf{v}_{j}^{(k)} \,,
% \end{aligned}\label{eqn:grad}
% \end{equation}
where $\text{MLP}$ represents a simple multilayer perceptron for time interval embedding $\mathbf{T}_{t_{ij}}$, and shares with different GNN layers.
Similar to GGNN, we feed the neighbor information $\mathbf{m}_i^{(k)}$ and the previous layer item $\mathbf{v}_i^{(k)}$ into \emph{GRU} to update the item-item relation node representations:  
\begin{equation}
\setlength{\abovedisplayskip}{3pt}
\setlength{\belowdisplayskip}{3pt}
\begin{aligned}
    \mathbf{v}_{i}^{v\xrightarrow{}v,(k+1)}=\text{GRU}\left(\mathbf{v}_{i}^{(k)},\mathbf{m}_{i}^{(k)} \right)\,,
\end{aligned}
\end{equation}
where the \emph{GRU unit} is parameters-shared for all item nodes updating at current layer.
With the combination of contextual interaction items representations and previous cross-semantic item presentations, we integrate the chronological item transition information into node embedding in the item-item relation.

\vspace{-0.15cm}
\subsubsection{Interest Extraction Layer.}\label{sec:i2v}
%1.兴趣节点的更新过程即为兴趣提取过程。
%2.我们基于注意力机制来学习每个兴趣节点与物品的关系，提取物品对抽象兴趣的贡献程度。
For the item-interest relation at layer $k$, a graph attention neural network is utilized for updating interest node embedding, as the process of distilling interest representation.
In particular, we compute the corresponding correlation weight $\alpha_{ij}^{(k)}$ between target interest node $u_i$ and neighbor item $v_j$, as the explainable assignment scores for disentangled interest.
And interest node representation is updated via the sum pooling as following:
\begin{equation}
\setlength{\abovedisplayskip}{3pt}
\setlength{\belowdisplayskip}{3pt}
\begin{aligned}
%\mathbf{a}^{(k)}_r
    \alpha_{ij}^{(k)}=\underset{j\in \mathcal{N}_{v\xrightarrow[]{}u}(i)}{\text{softmax}}\left(\text{LeakyReLU}\left( \mathbf{W}^{v\xrightarrow[]{}u,(k)}_{u} \mathbf{u}_{i}^{(k)}+ \mathbf{W}^{v\xrightarrow[]{}u,(k)}_{v} \mathbf{v}_{j}^{(k)}\right)\right)\,,
\end{aligned}
\end{equation}
\begin{equation}
\setlength{\abovedisplayskip}{3pt}
\setlength{\belowdisplayskip}{3pt}
\begin{aligned}
    u_{i}^{v\xrightarrow[]{}u,(k+1)}=\sum_{j\in \mathcal{N}_{v\xrightarrow[]{}u} (i)}{\alpha_{ij}^{(k)}} \mathbf{W}_{trans}^{v\xrightarrow[]{}u,(k)} \mathbf{v}_{j}^{(k)}\,.
\end{aligned}
\end{equation}
where $\mathbf{W}_{trans}^{v\xrightarrow[]{}u,(k)} \in \mathbb{R}^{d\times d}$ represents the information transformation matrix of neighbor nodes, $\mathbf{W}_{u}^{v\xrightarrow[]{}u,(k)}, \mathbf{W}_{v}^{v\xrightarrow[]{}u,(k)} \in \mathbb{R}^{1\times d}$ are the parameters of linear transformation for the target interest and the source item, respectively.

Meanwhile, we also update the center timestamp and temporal compactness of each interest node based on the normalized correlation coefficient:
\begin{equation}
\setlength{\abovedisplayskip}{3pt}
\setlength{\belowdisplayskip}{3pt}
\begin{aligned}
    t_{cent,i}^{(k+1)}=\sum_{j\in \mathcal{N}_{v\xrightarrow[]{}u} (i)} \alpha_{ij}^{(k)} t_{j,1}  \;\;;\;\; 
    t_{comp,i}^{(k+1)}=\sum_{j \in \mathcal{N}_{v\xrightarrow[]{}u} (i)} \alpha_{ij}^{{k}} |t_{j,1}-t_{cent,i}^{(k)}|.
\end{aligned}
\end{equation}

Through the assignment score $\alpha_{ij}^{(k)}$, we not only distill the latent representations of multi-view interests from the sessions with overlapped and interwove interests, but also offer an explanations parameter for each interest factor.

\vspace{-0.15cm}
\subsubsection{Interest Attaching Layer}\label{sec:v2i}
To refine the item representation via the disentangled interest, we adopt a graph attention network to update the item representation under the interest-item semantic.
At first, we estimate the attention coefficient between target item and source interest node.
Here, we consider both the similarity and  temporal continuity of each item and interest node pair:
\begin{equation}
\setlength{\abovedisplayskip}{3pt}
\setlength{\belowdisplayskip}{3pt}
\begin{aligned}
    \beta_{ij}^{(k)}=\left(\mathbf{W}^{u\xrightarrow[]{}v,(k)}_{v} \mathbf{v}_{i}^{(k)}\right)^T \mathbf{W}^{u\xrightarrow[]{}v,(k)}_{u} \mathbf{u}_{j}^{(k)}+\mathbf{w}_t^T \mathbf{T}_{t_{ij}^{u\xrightarrow[]{}v,(k)}} +{b}_t,
\end{aligned}
\end{equation}
\begin{equation}
\setlength{\abovedisplayskip}{3pt}
\setlength{\belowdisplayskip}{3pt}
\begin{aligned}
    \beta_{ij}^{(k)}=\underset{j\in \mathcal{N}_{u\xrightarrow[]{}v}(i)}{\text{softmax}}\left(\mathrm{LeakyReLU}\left(\beta_{ij}^{(k)} \right)\right)\,,
\end{aligned}
\end{equation}
%continuity
where $\mathbf{W}^{u\xrightarrow[]{}v,(k)}_{v}, \mathbf{W}^{u\xrightarrow[]{}v,(k)}_{u} \in \mathbb{R}^{d\times d}$ are the transforming matrices for the item and interest.
%, $\sigma$ is an activate function
% $\mathbf{t}_{ij}^{u\xrightarrow[]{}v,(k)}$ is the temporal embedding for 
$t_{ij}^{u\xrightarrow[]{}v,(k)}=\frac{\left|t_{cent,j}^{(k)}-t_{i,1}\right|}{t_{comp,j}^{(k)}}$ denotes the distance between item and interest in timeline, and $ \mathbf{w}_t \in \mathbb{R}^{d},  b_t$ are the linear transformation of it.
Here, we utilize the residual between the center timestep of interest and timestep of items regularized by the interest compactness value, to measure the similarity of interest and item pair in the temporal view.

Then, gathered interest factors are scattered to item nodes for generating the intention-augmented item representations via parameters  $\mathbf{W}_{trans}^{u\xrightarrow[]{}v,(k)} \in \mathbb{R}^{d\times d}$:
\begin{equation}
\setlength{\abovedisplayskip}{3pt}
\setlength{\belowdisplayskip}{3pt}
\begin{aligned}
    \mathbf{v}_{i}^{u\xrightarrow[]{}v,(k+1)}=\sum_{j\in \mathcal{N}_{u\xrightarrow[]{}v} (i)}{\beta_{ij}^{(k)}} \mathbf{W}_{trans}^{u\xrightarrow[]{}v,(k)} \mathbf{u}_{j}^{(k)}\,,
\end{aligned}
\end{equation}

The edges between interests and items indirectly connect each item to all other items through the intermediate interest node.
Compare to the original sparse item-item graph, the additional interest nodes and edges help GNNs to effectively capture long-range dependencies in sessions with any length because it propagates information along intermediate interest paths within two-hop.

\vspace{-0.15cm}
\subsubsection{Layer Combination}
By updating the node representation based on diverse semantic relations synchronously, we obtain the item representations from adjacent items and related interest, and the interest embeddings based on item-level soft clustering.
Then, we gather cross-semantic information through average operation like RGCN \cite{rgcn}, i.e., $\mathbf{v}_{j}^{(k+1)}=\sigmoid(\mathbf{v}_{j}^{u\xrightarrow[]{}v,(k+1)}+\mathbf{v}_{j}^{v\xrightarrow[]{}v,(k+1)})/2$,  $\mathbf{u}_i^{(k+1)}=\sigmoid(\mathbf{u}_i^{v\xrightarrow[]{}u,(k+1)})$, and consider $\mathbf{v}_j^{(k+1)}$ and  $\mathbf{u}_i^{(k+1)}$ as the output of item node $v_j$ and interest node $u_i$ after $k$-th GNN layers.

Moreover, in order to mine deeper items transition relations, multi-layers of GNN are stacked to propagate high-order information.
Moreover, we utilize the gated mechanism to balance the item node representations between initial embedding and $K$ layers output, as follows:
\begin{equation}
\setlength{\abovedisplayskip}{3pt}
\setlength{\belowdisplayskip}{3pt}
\begin{aligned}
    g=\sigmoid(\mathbf{W}_{gated}\left[\mathbf{v}_{j}^{(0)} \parallel \mathbf{v}_{j}^{(K+1)}\right])\,,
\end{aligned}
\end{equation}
\begin{equation}
\setlength{\abovedisplayskip}{3pt}
\setlength{\belowdisplayskip}{3pt}
\begin{aligned}
    \mathbf{v}_{j}=g \mathbf{v}_{j}^{(0)} + (1 - g) \mathbf{v}_{j}^{(K+1)} \,.
\end{aligned}
\end{equation}
where $\parallel$ is the concatenate operation,  $\mathbf{W}_{gated} \in \mathbb{R}^{2d}$ and sigmoid function $\sigmoid(\cdot)$ generate the balance factor $g$ to alleviate over-smooth problems of deep GNN.
As for the interest node, we simply adopt the multi-layers'  output as the final  latent interest representation, i.e., $\mathbf{u}_i=\mathbf{u}_i^{(K+1)}$.

% Moreover, each disentangled representation is also associated with its explanatory graphs to explicitly present the intents, i.e., the weighted adjacency matrix. Such explanatory graphs are able to show reasonable evidences of what information construct the disentangled representations.
\vspace{-0.1cm}
\subsection{Session Representation Learning Layer}
\vspace{-0.05cm}
% To extract final session representations based on the temporal-attached multi-interest graph, the aggregating representation of items in that session based on each abstract interest factors requires comprehensive consideration. 
After the Disentangle Graph Modeling, we obtain the final item and interest node embeddings.
Then we aggregate the representation of items in the session based on each abstract interest factor to generate the session representations.
% with additional last-item based time interval information
Moreover, the items clicked later in the session usually reveal the user’s current intentions better, which draws greater attention for SBR.
Therefore, we incorporate additional last-item based time interval information into interest-based attention to capture user activate intention dynamically.
% After feeding a session sequence into Disentangle GNN, we obtain the representation of the interests and items involved in the session, i.e., $ $. 

In detail, inspired by the reverse position embedding in GCE-GNN\cite{gce-gnn}, we integrated the last-item based time interval information with the obtained item representations, which extends the position information to a more fine-grained time domain to make a better prediction:
\begin{equation}
\setlength{\abovedisplayskip}{3pt}
\setlength{\belowdisplayskip}{3pt}
\begin{aligned}
    \mathbf{z}_{i}=\text{tanh}\left( \mathbf{W}_0\left[\mathbf{v}_{i} \parallel \mathbf{T}_{t_{L,i}}\right]+\mathbf{b}_0\right)\,,
\end{aligned}
\end{equation}
where $\mathbf{W}_0 \in \mathbb{R}^{d\times 2d}$ and $\mathbf{b}_0 \in \mathbb{R}^{d} $ are trainable parameters, $\mathbf{T}_{t_{L,i}} $ is the embedding for last-time based interval $t_{L,i}$.
% Hence we incorporate last-item based time interval information and session information to make prediction.
% through concatenation and non-linear transformation operation,
% and $l$ is the length of the current interaction sequence.

% Meanwhile, we consider coarse-level interest fusion representation by average pooling over all disentangled latent interests in this session: 
% \begin{equation}
% \setlength{\abovedisplayskip}{3pt}
% \setlength{\belowdisplayskip}{3pt}
% \begin{aligned}
%     \mathbf{s}_{I}=\frac{1}{H}\sum_{i=1}^{H} \mathbf{u}_i^{(K)}\,,
% \end{aligned}
% \end{equation}
% where $ H $ is the pre-defined hyper-parameter of the number of interest node.
Then, the intentions of user are learnt by a shared attention mechanism, which dynamically weights item representation based on each interest $\mathbf{u}_{h}$:
\begin{equation}
\setlength{\abovedisplayskip}{3pt}
\setlength{\belowdisplayskip}{3pt}
\begin{aligned}
    \gamma_{i,h}=\mathbf{q}^T\sigmoid(\mathbf{W}_1\mathbf{z}_{i}+\mathbf{W}_2\mathbf{u}_{h}+\mathbf{b})\,,\;\;
    \mathbf{s}_{G,h}=\sum_{i=1}^{N}\gamma_{i,h}\mathbf{v}_{i}\,,
\end{aligned}
\end{equation}
where $\sigmoid$ is an activate function, $\mathbf{W}_1\,, \mathbf{W}_2 \in \mathbb{R}^{d \times d}$ and $\mathbf{q}\,,\mathbf{b}\in \mathbb{R}^d$ are trainable parameters. 
Then we combine the $h$-th coarse-level intention and refined item-augment representations to generate the session representation for each interest:
\begin{equation}
\setlength{\abovedisplayskip}{3pt}
\setlength{\belowdisplayskip}{3pt}
\begin{aligned}
    \mathbf{S}_{h}=\sigmoid\left(\mathbf{W}_3\left[\mathbf{s}_{G,h} \parallel \mathbf{u}_{h}\right]\right)\,,
\end{aligned}
\end{equation}
where $\mathbf{W}_3 \in \mathbb{R}^{d \times 2d}$ is the trainable parameter.

With the incorporation of auxiliary last-item based interval information, we capture session representations involved in both the disentangled interest factors and interaction session items in relative chronological temporal-aware patterns.

\vspace{-0.1cm}
\subsection{Prediction and Training}\label{sec:prediction}
\vspace{-0.05cm}
Based on each disentangled interest representation $\mathbf{S}_h$ learnt above and the normalized initial embeddings $\mathbf{v}_{i}'$ of candidate items, we then estimate the interaction  probability $\hat{\mathbf{y}}$ of candidate items for current session:
\begin{equation}
\setlength{\abovedisplayskip}{3pt}
\setlength{\belowdisplayskip}{3pt}
\begin{aligned}
    \hat{\mathbf{y}}_{i}=\max_{1\leq h \leq H} \text{softmax}\left(\mathbf{S}_h^T \mathbf{v}_{i}' \right)\,, \;\;
    \mathbf{v}_{i}' = \mathbf{v}_{i}^{(0)}/{\parallel\mathbf{v}_{i}^{(0)}\parallel_2}
\end{aligned}
\end{equation}
%\mathbf{v}_{i}' = \frac{\mathbf{v}_{i}^{(0)}}{\left\|\mathbf{v}_{i}^{(0)}\right\|}
where $\hat{{y}}_i \in \hat{\mathbf{y}}$ denotes the probability that the user will click on item $v_i$ in the current session, and $H$ is the pre-defined parameter of the interest node number.

As mentioned above, flexible number of interest nodes encourages the chunked interest representations conditioned on different behavior patterns.
However, the difference constraint drove by multiple interest extractions is insufficient: there might be redundancy among latent interests representation, which conflicts with the target of disentangling multi-view user interest.
% e.g., if one factor interest representation $u_i$ can be inferred by other interests $\{u_{j}|j\neq{i}\}$, the interest cluster $i$ is highly likely to  overlap and interweave with other interests,
We hence introduce interest independents loss, which hires distance correlation measures as a regularizer, with the target of encouraging the multi-interest representations to be diverse.
We formulate this as follows:
% \begin{equation}
% \setlength{\abovedisplayskip}{3pt}
% \setlength{\belowdisplayskip}{3pt}
% \begin{aligned}                                 
%     k_{ij}=\sum_{i=1}^{H}\sum_{j=i+1}^{H}\text{cos}\left(\mathbf{u}_i,\mathbf{u}_j\right)\,,
% \end{aligned}\label{eqn:loss2}
% \end{equation}
% \begin{equation}
% \setlength{\abovedisplayskip}{3pt}
% \setlength{\belowdisplayskip}{3pt}
% \begin{aligned}                                
%     \mathcal{L}_{dis}=\sum_{i=1}^{H}\sum_{j=i+1}^{H}l_{ij}\,,
% \end{aligned}\label{eqn:loss2}
% \end{equation}
\begin{equation}
\setlength{\abovedisplayskip}{3pt}
\setlength{\belowdisplayskip}{3pt}
\begin{aligned}                                
    \mathcal{L}_{corr}= \sum_{i=1}^{H}\sum_{j=i+1}^{H}\text{cos}\left(\mathbf{u}_i,\mathbf{u}_j\right)\,,
\end{aligned}\label{eqn:loss2}
\end{equation}
where $cos\left(\cdot\right)$ indicates the similarity distance between two inner-session interest representation pair.

The final optimization process is to minimize the cross entropy loss function together and the interest-independence loss jointly:
\begin{equation}
\setlength{\abovedisplayskip}{3pt}
\setlength{\belowdisplayskip}{3pt}
\begin{aligned}                                 
    \mathcal{L}=-\sum_{i=1}^{|\mathcal{V}|}\mathbf{y}_i\text{log}(\hat{\mathbf{y}}_i)+(1-\textbf{y}_i)\text{log}(1-\hat{\mathbf{y}}_i)+\lambda\mathcal{L}_{corr}\,,
\end{aligned}
\end{equation}
where $\mathbf{y}_i \in \mathbb{R}^{|\mathcal{V}|}$ is a one-hot vector of ground truth, and $\lambda $ is the coefficient controlling interest-independence term.

%baseline:
% gru4rec,NARM,RIB(?)
% SRGNN,StarGNN,LESSR
% NARM+,SRGNN+,LESSR+ (加入时间信息)
% 
% 其他：兴趣分解的那几篇 (DIN,SUPGE,TASER) X

%灼烧实验：
% 完全无时间,有部分时间（3个版本起步）
% 没有autohead、直接指定head数
% 没有head之间的差异loss
% 构图灼烧

%超参实验：
% 不同的max head数
% 长短session的performance比较（此处可以和其他模型一起对比，并结合不同的max head进行对比）
% 不同的GNN层数的比较（此处可以和StarGNN对比，它应该是是层数上升、效果下降的）
% 不同的时间片长度（time unit）

\section{Experiments}

In this section, we conduct experiments on SBR to evaluate the performance of our method compared with other state-of-the-art models  \footnote{Our code and data will be released for research purpose.}. Our purpose is to answer the following research questions:$\;$
\textbf{RQ1:} How does our model perform compared with state-of-the-art SBR methods?  $\;$
\textbf{RQ2:} How does the temporal information of the sequence affect the recommendation results?$\;$
\textbf{RQ3:} Is the design of our model reasonable and effective? How do the key modules of TMI-GNN influence the model performance?$\;$
\textbf{RQ4:} How do the hyper-parameters affect the effectiveness of our model?

\newcommand{\tabincell}[2]{\begin{tabular}{@{}#1@{}}#2\end{tabular}}  

\begin{table}[htbp]
\vspace{-0.2cm}
    \caption{Statistics of datasets used in experiments.}
    \label{tab:dataset}
    \centering
    \small
    \begin{tabular}{lrrr}
    \toprule
    Statistic& RetailRocket & Yoochoose & Jdata \\
    \midrule
    No. of items  & 45,630 & 19,589  & 92,792 \\
    No. of sessions  & 341,831 & 975,060 & 3,397,208 \\
    Avg. of session length & 4.91 & 6.87 &  6.68\\
    Avg. of time interval (seconds) &65.8  &51.3 &52.6 \\
    % \tabincell{l}{Time interval (seconds)\\
    % (25\%/50\%/75\%)} &26/65/163  &19/51/115 &21/52/131  \\
    No. of train sessions & 314,924  & 777,025  & 2,862,842 \\
    No. of test sessions & 26,907 & 198,035  & 534,366 \\
    
    \bottomrule
    \end{tabular}
    \vspace{-0.2cm}
\end{table}

\begin{table*}[t]
    \centering
    \caption{Experimental results (\%) of different models in H@\{10, 20\}, and N@\{10, 20\} on three datasets. $Improv.$ means improvement over the state-of-art methods.The bold number indicates the improvements over the best baseline (underlined) are statistically significant ($p \textless 0.01$) with paired t-tests.}
    %  The * means the best results on baseline methods.
    % $Improv.$ means improvement over the state-of-art methods.
    \label{tab:overall}
        % \resizebox{\textwidth}{30mm}{
    \begin{tabular}{p{1.4cm}<{\centering}p{0.85cm}<{\centering}p{0.85cm}<{\centering}p{0.95cm}<{\centering}p{1.0cm}<{\centering}p{0.05cm}p{0.85cm}<{\centering}p{0.85cm}<{\centering}p{0.95cm}<{\centering}p{1.0cm}<{\centering}p{0.05cm}p{0.85cm}<{\centering}p{0.85cm}<{\centering}p{0.95cm}<{\centering}p{1.0cm}<{\centering}}%{lrrrrrrrrrrrr}
    \toprule
    \multirow{2}{*}{ \bfseries Models}& \multicolumn{4}{c}{ \bfseries RetailRocket }& & \multicolumn{4}{c}{\bfseries Yoochoose 1/16 }& &\multicolumn{4}{c}{\bfseries Jdata} \\
    \cline{2-5}
    \cline{7-10}
    \cline{12-15}
    &H@10 &H@20 &N@10 &N@20 && H@10& H@20 &N@10 &N@20 && H@10 &H@20 &N@10 &N@20\\
   % \cline{1-7}
    \midrule
    %\cline{1-7}
    GRU4Rec &29.40 &37.32 &18.64 &20.73  && 44.91 &54.12  &28.16  &30.29  && 28.62 &39.05  &15.67  &18.58 \\
    %\cline{1-7}
    NARM &32.96 &\underline{41.09} &\underline{20.23} &21.89  &&52.86  &63.12  &32.03  &34.64  &&31.47  &43.32  &16.41  &19.40   \\
    
    %\cline{1-7}
    SRGNN &32.92 &40.60 &20.13 &22.10  &&54.87  &64.63  &33.90  &36.38  &&33.01  &44.27  &18.88  &21.73 \\
    
    %\cline{1-7}
    LESSR &31.82 &40.48 &19.90 &22.09  &&55.51  &65.28  &34.51  &36.99  &&33.13  &44.01  &18.99  &21.74  \\
    %StarGNN
    DATMDI &32.26 &40.51 &20.06 &22.15 
    &&55.77  &65.58  &34.61  &37.12  &&33.24  &44.39  &19.05  &21.82  \\

    \midrule
    
    RIB &32.76 &40.96 &20.17 &21.78  &&52.80  &63.08  &31.97  &34.59  &&31.36  &43.24  &16.31  &19.33 \\
    
    $\text{SRGNN}^\dagger$ &\underline{32.97} &40.63 &20.11 &22.12  &&54.85  &64.64  &33.87  &36.39  &&33.10  &44.37  &18.93  &21.79 \\
    
    $\text{LESSR}^\dagger$ &31.90 &40.56 &19.94 &\underline{22.15}  &&55.59  &65.35  &34.60  &37.09  &&33.23  &44.12  &19.03  &21.76 \\
    
    $\text{DATMDI}^\dagger$  &32.46 &40.64 &20.12 &22.13 
    &&\underline{55.86}  &\underline{65.62}  &\underline{34.67}  &\underline{37.25}  &&\underline{33.29}  &\underline{44.46}  &\underline{19.09}  &\underline{21.88} \\
    
    %\cline{1-7}
    \midrule

    Ours &{\bfseries 33.07} &{\bfseries 42.36}& {\bfseries 20.31} &{\bfseries  22.20}& &{\bfseries 56.56}&{\bfseries 66.46} &{\bfseries 35.53}&{\bfseries 38.05} & &{\bfseries  33.77}&{\bfseries 45.38} &{\bfseries  19.21}&{\bfseries 22.18}\\
    \midrule
    {$Improv.$}& 0.30\% & 3.09\% & 0.39\% & 0.22\% & & 1.29\% & 1.30\% & 2.60\% & 2.48\% & & 1.47\% & 2.05\% & 0.79\% & 1.51\%    \\
    \bottomrule
    \end{tabular}
    % \vspace{-0.2cm}
\end{table*}

\vspace{-0.15cm}
\subsection{Experimental Setup}
\vspace{-0.1cm}
\subsubsection{Dataset.} 
We conduct extensive experiments on three public datasets: \emph{RetailRocket}, \emph{Yoochoose} and \emph{Jdata}, which are widely used in the SBR research \cite{lessr,gce-gnn,MKMSR} and can support our work with seconds-level timestamp information.
%数据集满足我们的实验需求

\begin{itemize}[leftmargin=*]
% \begin{inparaitem}
%
    \item \emph{RetailRocket}\footnote{https://www.kaggle.com/retailrocket/ecommerce-dataset} contains behavior data and item properties that collected from a real-world e-commerce website.
    \item \emph{Yoochoose}\footnote{http://2015.recsyschallenge.com/challege.html} is a RecSys Challenge dataset, which consists of click-streams from an E-commerce website. Different from \cite{liu2018stamp,narm}, we use the most recent fractions 1/16 sequences of Yoochoose as the total dataset \emph{Yoochoose} 1/16.
    \item \emph{Jdata} \footnote{https://jdata.jd.com/html/detail.html?id=8} records the historical interactions from the JD.com. It contains a stream of user actions within two month. We extract the session data with the setting of the duration time threshold 1 hour.
\end{itemize}

To filter poorly informative sessions and items, following \cite{narm,srgnn}, we first filtered out all sessions of length $\leq2$ and items appearing less than 5 times in all datasets.
Then we applied a data augmentation technique described in \cite{narm}.
The statistics of all datasets after prepossessing are summarized in \autoref{tab:dataset}.

\vspace{-0.15cm}
\subsubsection{Baseline Models.}
To demonstrate TMI-GNN’s superiority performance, we compare it with several representative competitors, including the state-of-the-art SBR models and several temporal-concerned methods.

\begin{itemize}[leftmargin=*]
% \begin{inparaitem}
% \footnote{https://github.com/hidasib/GRU4Rec}
    \item \textbf{GRU4Rec}\cite{gru4rec} employs the GRU to capture the representation of the item sequence simply.
    % \footnote{https://github.com/lijingsdu/sessionRec\_NARM}
    \item \textbf{NARM} \cite{narm} is a RNN-based model which combines with attention mechanism to generate the session embedding.
    
    \item \textbf{RIB} \cite{zhou2018micro} is a framework using RNN and attention layer to model user micro behavior including behavior and dwell time. Here we consider the time intervals as the dwell time and ignore the behavior types.
    % \footnote{https://github.com/CRIPAC-DIG/SR-GNN}
    \item \textbf{SRGNN} \cite{srgnn} converts session sequences into directed unweighted graphs and utilizes a GGNN layer \cite{li2015gated} to learn the patterns of item transitions.
% \footnote{https://github.com/twchen/lessr}
    \item \textbf{LESSR} \cite{lessr} adds shortcut connections between items in the session and considers the sequence information in graph convolution by using GRU.
    % \footnote{https://github.com/Peiyance/StarGNN}
    % \item \textbf{StarGNN} \cite{stargnn} utilizes a star graph neural network to model the complex transition pattern in sessions.
    % apply a highway networks to dynamically select the item embeddings before and after StarGNN.
    
    \item \textbf{DATMDI} \cite{DATMDI}  combines the GNN and GRU to learn the cross-session enhanced session representation.
    % \item \textbf{GCE-GNN}\footnote{https://github.com/CCIIPLab/GCE-GNN} \citep{gce-gnn} aggregates the global context  and the item sequence in the current session to generate the session embedding through different level graph neural networks.
\end{itemize}

Besides, we combine the time interval embdding with ID embedding as input following \cite{zhou2018micro}, and inject additional time information into the graph by adding learnable time interval weights like \autoref{sec:i2i} for SRGNN, LESSR and DATMDI method separately, named $\text{SRGNN}^\dagger$,  $\text{LESSR}^\dagger$, $\text{DATMDI}^\dagger$.

\subsubsection{Evaluation Metrics.}
To evaluate the recommendation performance, we employ two widely used metrics: Hit ratio (H@$k$)  and Normalized discounted cumulative gain (N@$k$) following \cite{srgnn,gce-gnn}, where $k$ is 10 or 20. The average results over all test session are reported.

\vspace{-0.15cm}
\subsubsection{Implementation Details.}
We implement the proposed model based on Pytorch and DGL.
The embedding dimension is set to $128$.
All parameters are initialized through a Gaussian distribution with a mean of 0 and a standard deviation of $0.1$.
We employ the Adam optimizer to train the models with the mini-batch size of 512. We conduct the grid search over hyper-parameters as follows: learning rate in $\{0.001,0.01,0.1\}$, learning rate decay in $\{0.01,0.05,0.1,0.5\}$, learning rate decay step in $\{2,3,4\}$, controlling factor $\lambda$ in $\{1,3,10,30\}$.
The maximal time-step $m$ is set to $300$, which is large enough for all sessions.
To make the comparison fairer, we range the hyper-parameters of baseline methods with the same tuning scopes of our experiments.

% \end{inparaitem}
\vspace{-0.2cm}
\subsection{Overall Comparison (RQ1)}
\vspace{-0.05cm}
To demonstrate the overall performance of the proposed model, we compared it with the state-of-the-art methods for SBR. 
We can obtain the following significant observations from the comparison results shown in \autoref{tab:overall}.

\noindent \textbf{Comparison of Different Baselines.} 
%1.模型内部对比，包括rnn对比gnn、srgnn对比能捕捉到long distance的model（stargnn、Lessr，这个可以多提一点）、以及我们的模型对比其他的模型
%2.对比其他的模型的time，我们的模型的time能更有效的利用
%3.对比其他模型的单interest，我们模型能通过提取多interest embedding来disentanlgle user（单interest的局限性） intention，从而能更加深刻的capture用户兴趣表征
%RNN内部对比
%可能要增加神经模型对比传统模型
The NARM performs significantly better than GRU4Rec, which indicates the effectiveness of attention mechanism to capture the user’s main motivation.
%RNN与GNN对比
In the comparison between RNN-based models and GNN-based models for \emph{Yoochoose} and \emph{Jdata}, the GNN-based models generally outperform the RNN-based models, which verified the certain advantage of GNN for SBR.
% This verified that using GNN to model the transition relationship between items has certain advantages.
% GNN内部对比
Moreover, we notice that LESSR can outperform SRGNN in most cases.
This may be due to the special design in LESSR for capturing long-range dependency between items in a session, which contributes to better performance on some long sessions.
% Moreover, we notice that LESSR and StarGNN can outperform SRGNN in most cases.
% This may be due to the special design in LESSR and StarGNN for capturing long-range dependency between items in a session, which contributes to better performance on some long sessions.
Meanwhile, for dataset \emph{RetailRocket}, the simple model NARM outperforms other complex models because of the insufficient training caused by the small amount of data.

% LESSR propose a shortcut graph attention layer which can directly propagate information along shortcut connections, while StarGNN use a star node to represent the transition relationship among the items in the session
%添加时间信息的模型与RNN、GNN的对比
\noindent \textbf{Significance of Temporal Information.}
Then we turn to the temporal information attached methods.
The RNN-based model RIB achieves superior performance to GRU4Rec but performs slightly inferior than NARM,  because RIB employs a relatively simple attention instead of last-item based attention.
Compared to the original models, the temporal-enhanced methods all achieve better performance in some degrees, which indicates the significance of temporal information in SBR. 
Moreover, the adapted methods perform better on small dataset RetailRocket, which indicates that the auxiliary temporal information is more helpful for sparse interaction data.

\noindent \textbf{Model Effectiveness.}
% Then we move to the proposed model TMI-GNN.
We find that our model comprehensively outperforms all other baselines substantially on almost all metrics, which justifies the effectiveness of our model.
The performance improvement can be explained in two aspects.
One is that TMI-GNN can disentangle user intention via extracting multi-interests from the multi-interests session graph, so it can portray more profound representation of user interests.
This strategy break the limitation of expression ability of only one interest.
Another one is that we introduce multi-form temporal information to the process of disentangle graph modeling and session representation learning.
Compare with other time-aware models, TMI-GNN models diversified temporal information adequately effectively.

\begin{table}[htbp]
\vspace{-0.1cm}
    \caption{The performance comparison w.r.t different temporal information and module design.}
    %  w.r.t different information and different module design.
    \label{tab:information}
    \centering
    \small
    \setlength{\tabcolsep}{3pt}

\begin{tabular}{cl|cccc|cc|c}
\hline
\multicolumn{2}{c|}{\bfseries Model setting} &-V2V  &-U2V  &-Last &First  &-Interest &-Loss &Ours  \\ \hline
\multirow{2}{*}{\bfseries Yoochoose} &H@20  &66.32  &66.25  &66.28 &66.37 &65.86  &66.35  &66.46  \\ \cline{2-9}
                   & N@20 &37.54  &37.63  &37.53 &37.79  &37.23  &37.71  &38.05  \\ \hline
\multirow{2}{*}{\bfseries Jdata} &H@20  &44.89  &44.75  &44.82 &45.14 &44.25  &45.21  &45.38  \\ \cline{2-9}
                   & N@20 &21.98  &21.94  &21.97 &22.02 &21.80  &22.03  &22.18  \\ \hline
    % {NIN=3 } &66.46 &37.67 && 45.27 & 21.97    \\ 
    % {NIN=2} &66.44 &37.69 &&45.30 &22.09   \\
    % {NIN=1} &66.41 &37.98 &&45.34 &22.10   \\
\end{tabular}
    \vspace{-0.3cm}
\end{table}

\vspace{-0.1cm}
\subsection{Ablation Study (RQ2\&RQ3)}\label{sec:ablation}
\vspace{-0.05cm}
% In this subsection, we conduct some ablation studies on the proposed model to investigate the effectiveness of some designs. 

\noindent \textbf{Impact of temporal information.}
% \subsubsection{Impact of temporal information}
% In this part, we compare our model with several representative  baselines in  \autoref{tab:information} to test  whether considering  the  multiplex temporal sequence indeed boosts the performance of SBR.  
In this part, we compare our model with partially temporal information masked versions in \autoref{tab:information} to test whether considering the multi-form temporal information can boost model performance.
The method with "-V2V" means skipping the time intervals in item-level message propagation, "-U2V" indicates ignoring the temporal factors in interest-item relation, "-Last" represents removing the last-item based time-interval signals, and "First" means utilizing the first-item based time-interval embedding in session representation learning module.
% The method with “w/o T@V2V”, "w/o T@U2V" and "w/o T@Pooling", means skipping the time intervals in item-level information propagation, the  interest-item  relation temporal continuity factors, and  the temporal pattern signals in session representation learning module.
By comparing methods mentioned above, we find that the loss of any type of temporal information will cause the decline of model performance. 
Besides, compare to "First" time embedding, the last-item based time-interval information is more helpful for SBR.
Moreover, the loss of interest-item temporal continuity factors is more significant in our model compared with other temporal information types.
Based on the above illustrations, we demonstrate that injecting multi-form temporal information in our framework is indeed meaningful.

% \begin{table}[htbp]
% \vspace{-0.3cm}
%     \caption{The performance comparison w.r.t different information and module design. }
%     %  w.r.t different information and different module design.
%     \label{tab:information}
%     \centering
%     % \small
%     \setlength{\tabcolsep}{2pt}
%     \begin{tabular}{lrrrrrr}
%     \toprule
   
% \multirow{2}{*}{\bfseries Model setting} &
%   \multicolumn{2}{c}{Yoochoose} & &
%   \multicolumn{2}{c}{Jdata}  &
%   \\ \cline{2-3} \cline{5-6}  
%     & HR@20 &Nd@20 & &HR@20 &Nd@20  \\
%     \midrule
%     %\midrule
%     {w/o T@V2V} &66.32 &37.54 &&44.89 &22.02 \\
%     {w/o T@U2V}  &66.25 &37.63	&&44.75	&21.94	\\
%     {w/o T@Pooling}  &66.28 &37.53	&& 44.82	&22.97	\\
%     \midrule
%     {w/o Interest Node}&65.86 &37.23 && 44.25 & 21.80 \\
%     {w/o diverse loss}&66.35 &37.71 &&45.21 &22.03 \\
%     % {NIN=3 } &66.46 &37.67 && 45.27 & 21.97    \\  %interest node num adaptation
%     % {NIN=2} &66.44 &37.69 &&45.30 &22.09   \\
%     % {NIN=1} &66.41 &37.98 &&45.34 &22.10   \\
%     \midrule
%     Ours  &66.46 &38.05 &&45.38 &22.18 \\
%     \bottomrule
%     \end{tabular}
%     \vspace{-0.4cm}
% \end{table}

\begin{figure}[t]
    \centering
    \begin{subfigure}{0.43\linewidth}
        \includegraphics[width=\textwidth]{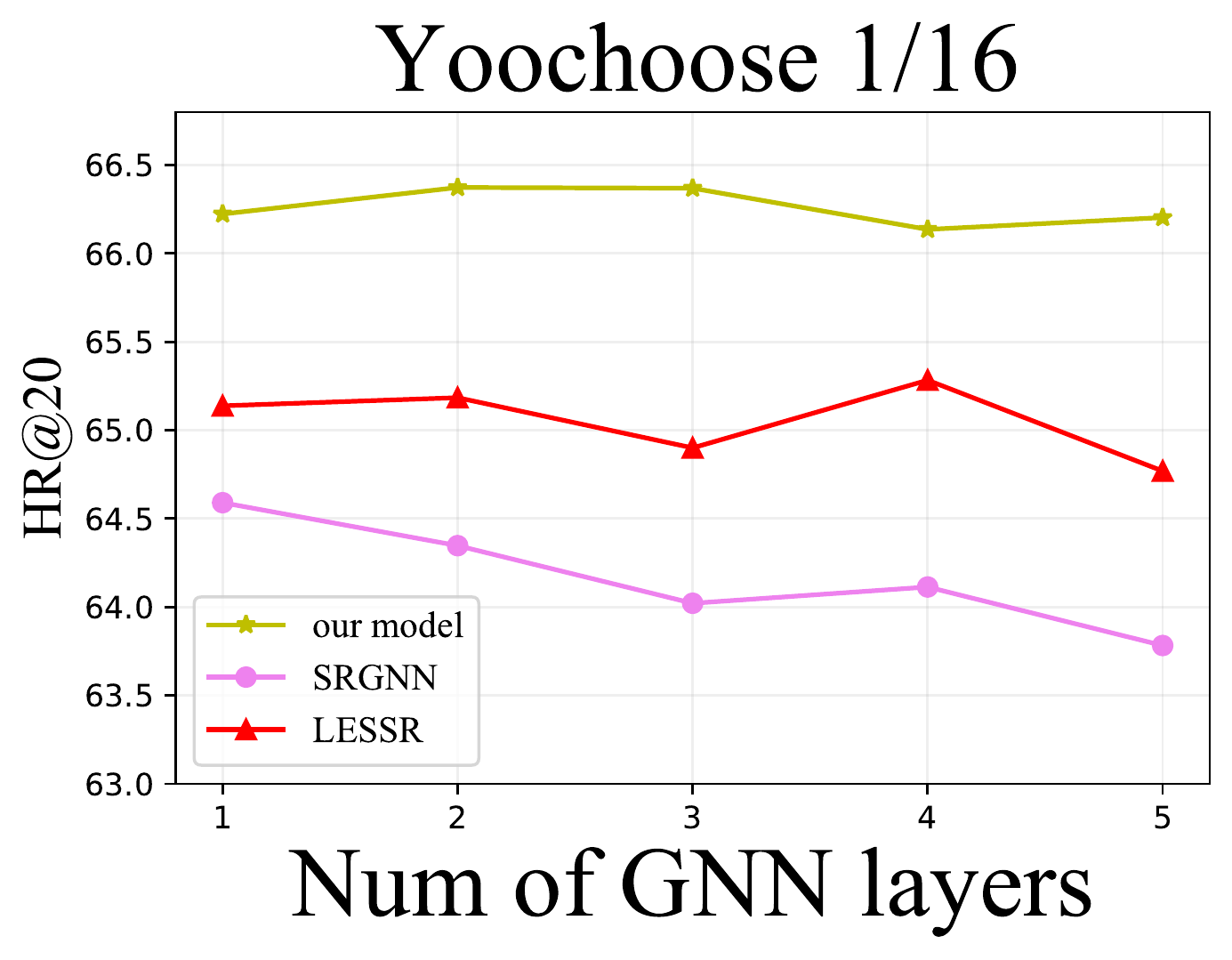}
    \end{subfigure}
    \begin{subfigure}{0.43\linewidth}
        \includegraphics[width=\textwidth]{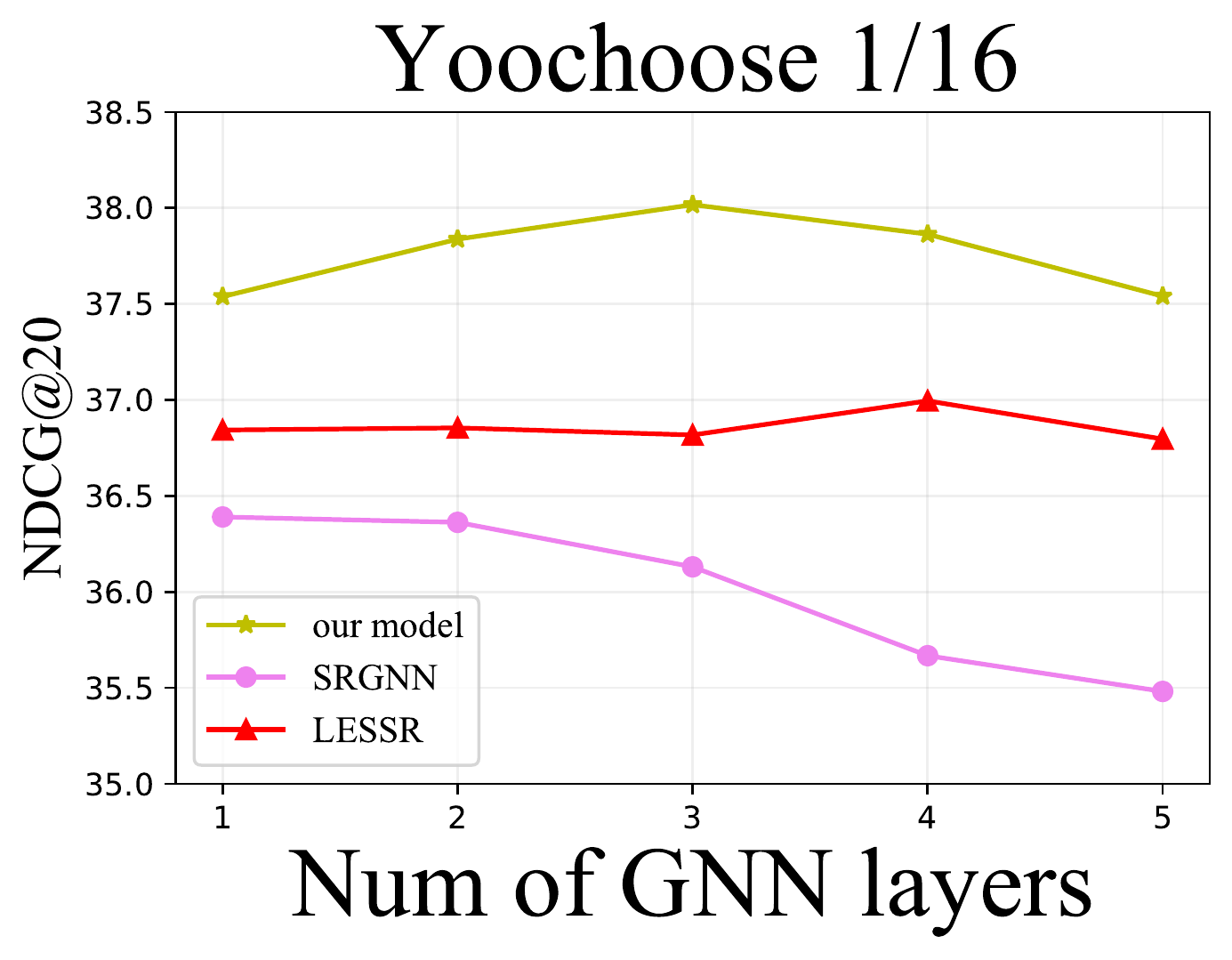}
    \end{subfigure}
    \begin{subfigure}{0.43\linewidth}
        \includegraphics[width=\textwidth]{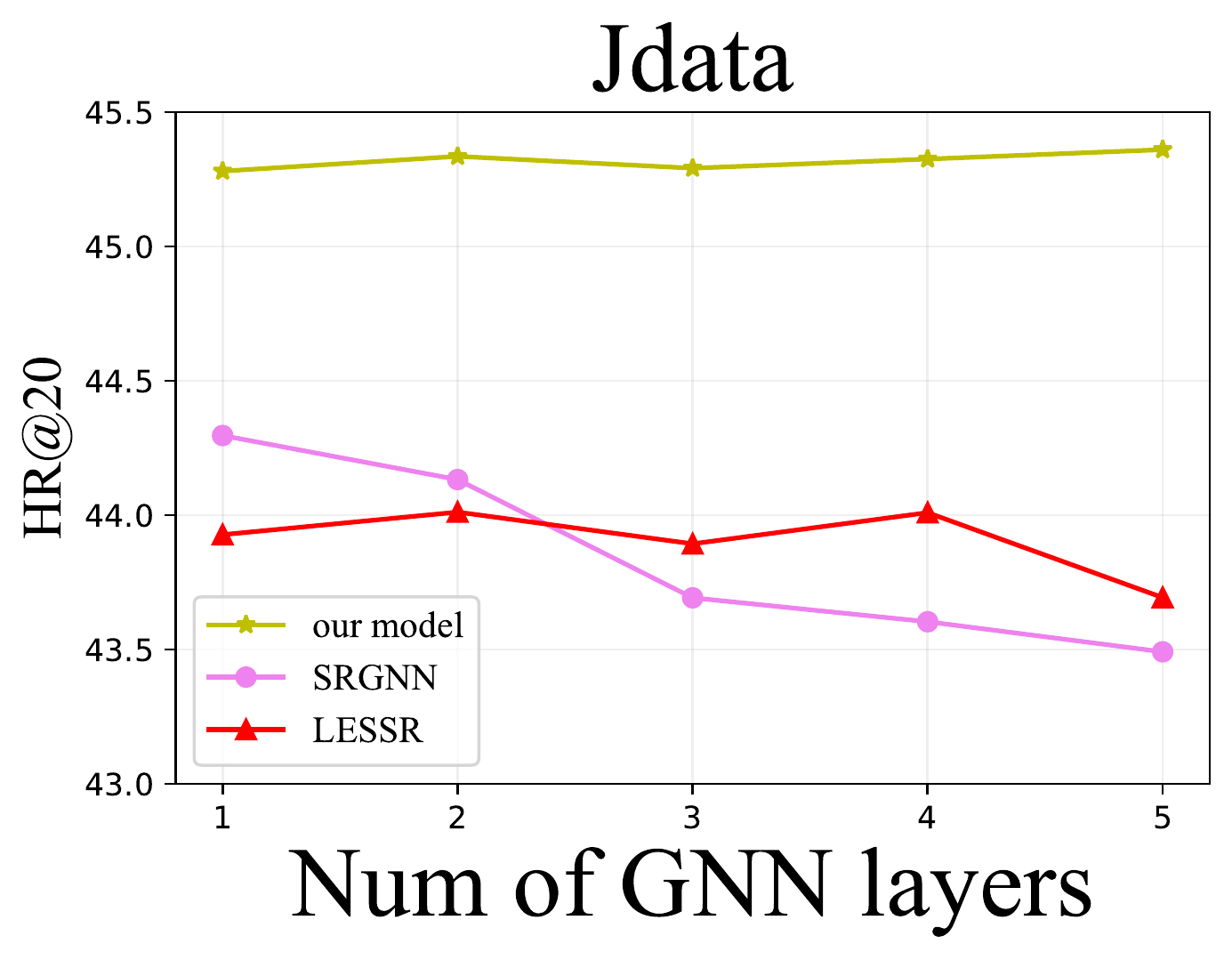}
    \end{subfigure}
    \begin{subfigure}{0.43\linewidth}
        \includegraphics[width=\textwidth]{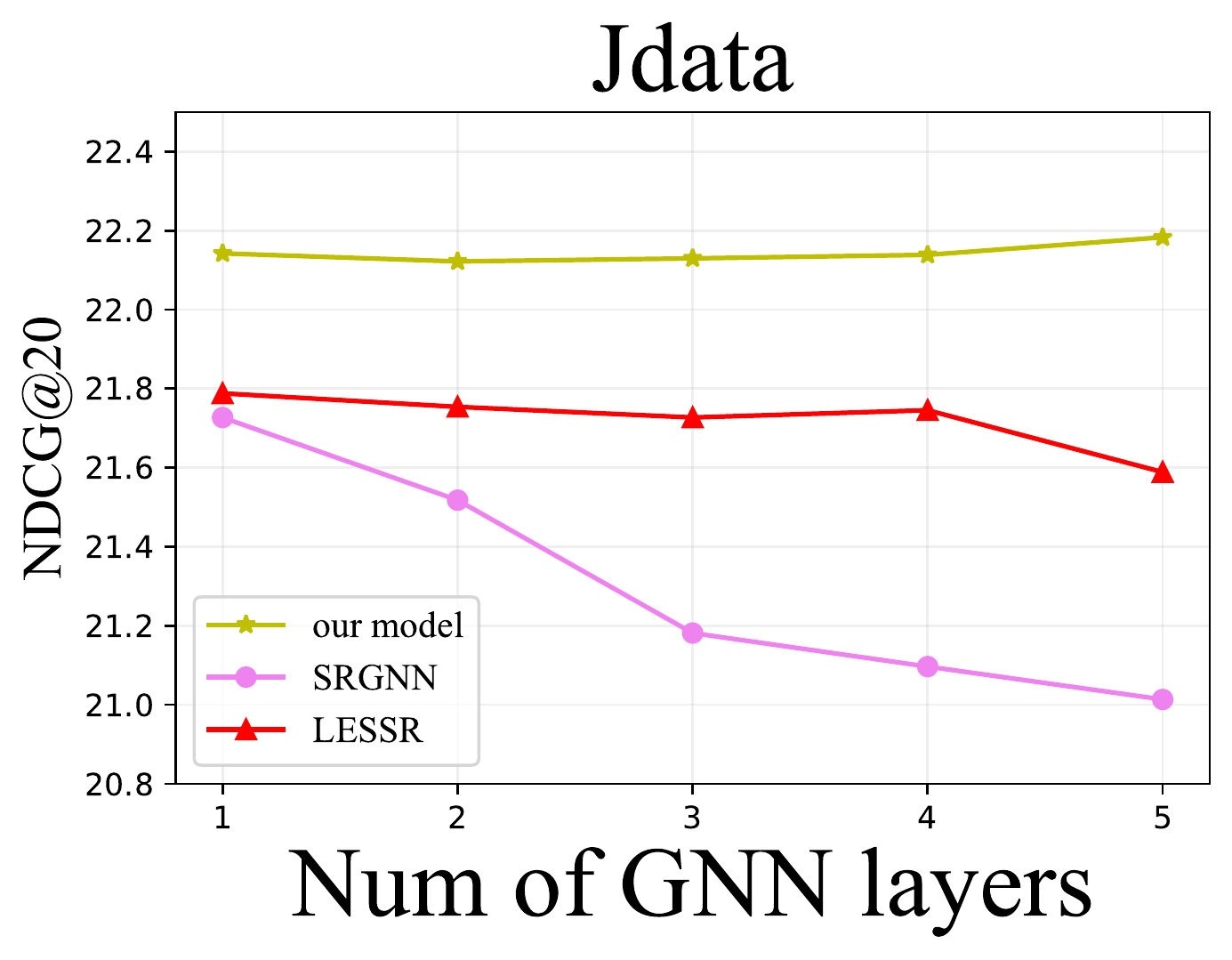}
    \end{subfigure}
    \caption{Performance comparison w.r.t. different depths of GNN.}
    \label{fig:hyper-depth-compare}
    \vspace{-0.4cm}
\end{figure}

\noindent \textbf{Impact of different Designs.}
% \vspace{-0.1cm}
% \subsubsection{Impact of different Layers }
In this part, we compare our method with different variants to verify the effectiveness of the critical components of TMI-GNN. 
Specifically, we remove the additional interest node in session graph (denote as ”-Interest”), and mask the interest independent loss (denote as ”-Loss”), respectively.
% Moreover, for ablating the influence of fixing the  number of interest node (NIN) , we also replace the \emph{Interest Node Num Adaption} module with interest node constant $\{1,2,3\}$ overall sessions.
The experimental results are presented in \autoref{tab:information}. 
It can be observed that the abstract interest nodes is pivotal for the model capability of intention capturing.
Meanwhile, for the diverse intention modeling, the removal of interest independent loss leads to great impact on model results, which demonstrates that the forced cross-interest separation is helpful for disentangling multiple interest of user.
% In addition, the fixed pattern of interest node number in multi-interest graph leads to model performance slightly drops, which further supports the necessarily of learnable interest node number.
In summary, we can infer that the key components of TMI-GNN are effective through the comparison and analysis above.

\vspace{-0.1cm}
\subsection{Hyper-parameters Study (RQ4) }\label{sec:hyper-res}
\vspace{-0.05cm}
% In this part, we perform experiments to explore how the hyper-parameters, such as the GNN depth and the time bucket width, influence the model performance.

\noindent \textbf{Impact of GNN depth.}
% \subsubsection{Impact of GNN Depth}
To study the impact of depth of GNN, we conducted experiments under different GNN depth settings (from 1 to 5).
Figure \ref{fig:hyper-depth-compare} shows the corresponding results.
We can observe that as the depth goes up from 1 to 3, the performance of our model increases on both datasets, which clarifies the significance of using multi-layer GNN to distill user's multi-interests.
Moreover, the performance declines as the depth grows from 3 to 5 on \emph{Yoochoose}, showing that excessive depth of GNN may lead to over-smoothing problem and less distinguishable item representation. 
As a comparison, with the number of GNN layers increasing, the performance of SRGNN drops on both two datasets, while our model's performance degrades slowly on \emph{Yoochoose}, even stay a  slightly increase on larger dataset \emph{Jdata}.
This indicates that our model is more efficient in handling the over-smoothing problem of GNN through the recombination of item representation, thus achieves more refined modeling of user interest through stacking more GNN layers without performance degradation.

\noindent \textbf{Impact of time bucket widths.}
% \vspace{-0.1cm}
% \subsubsection{Impact of Time Bucket Widths}
\begin{figure}[t]
\vspace{-0.1cm}
    \centering
    \begin{subfigure}{0.43\linewidth}
        \includegraphics[width=\textwidth]{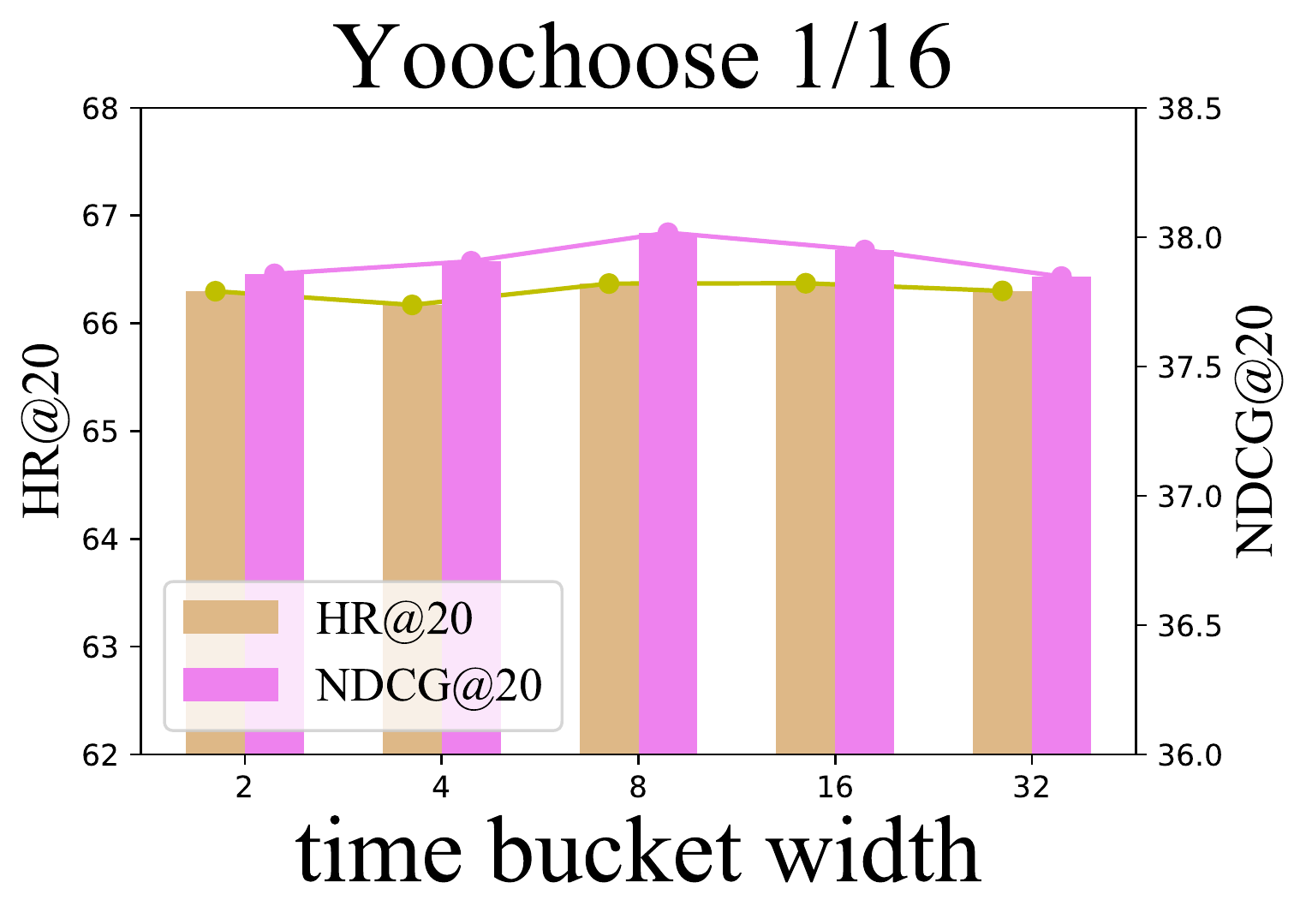}
    \end{subfigure}
    \begin{subfigure}{0.43\linewidth}
        \includegraphics[width=\textwidth]{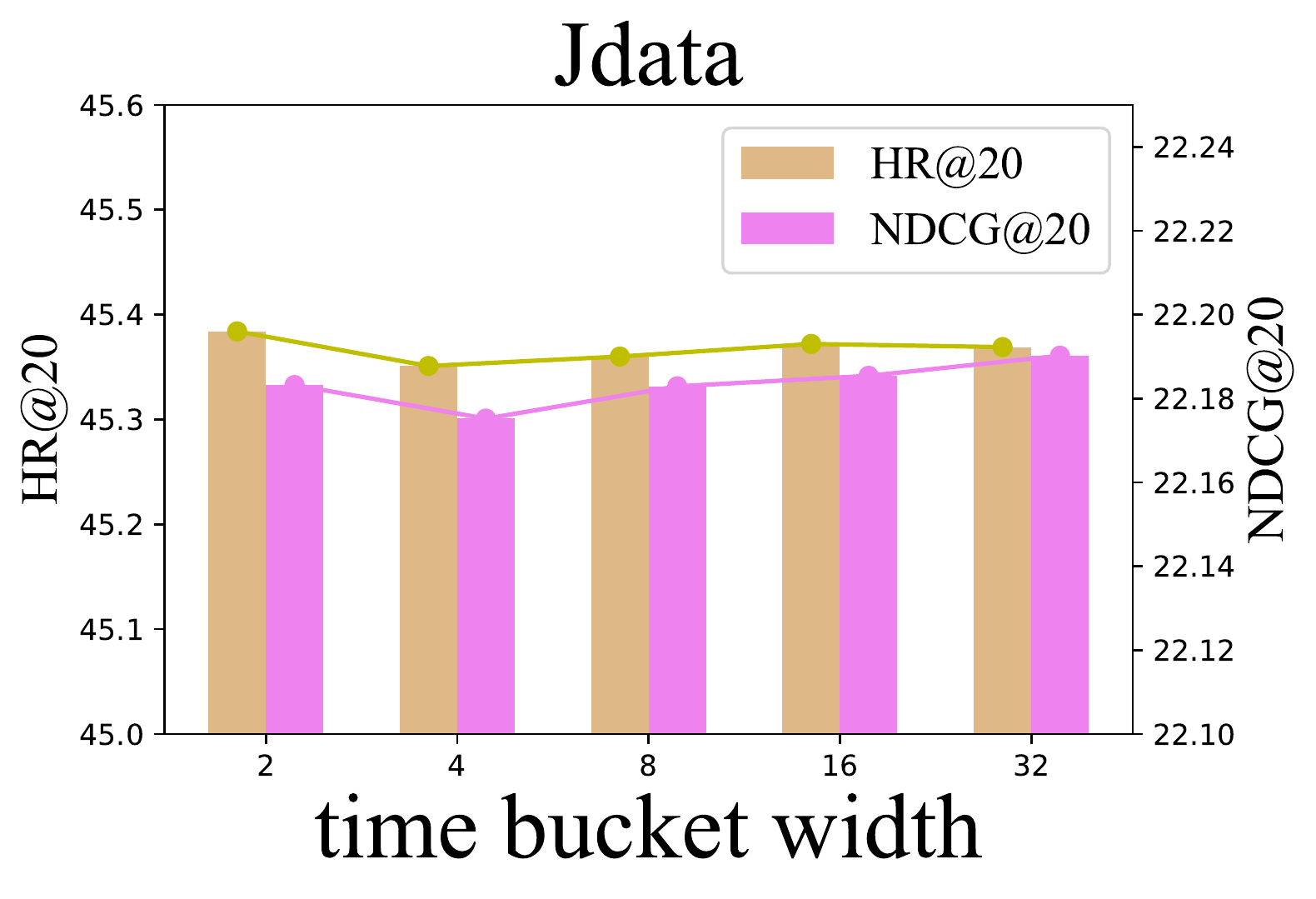}
    \end{subfigure}
    \begin{subfigure}{0.43\linewidth}
        \includegraphics[width=\textwidth]{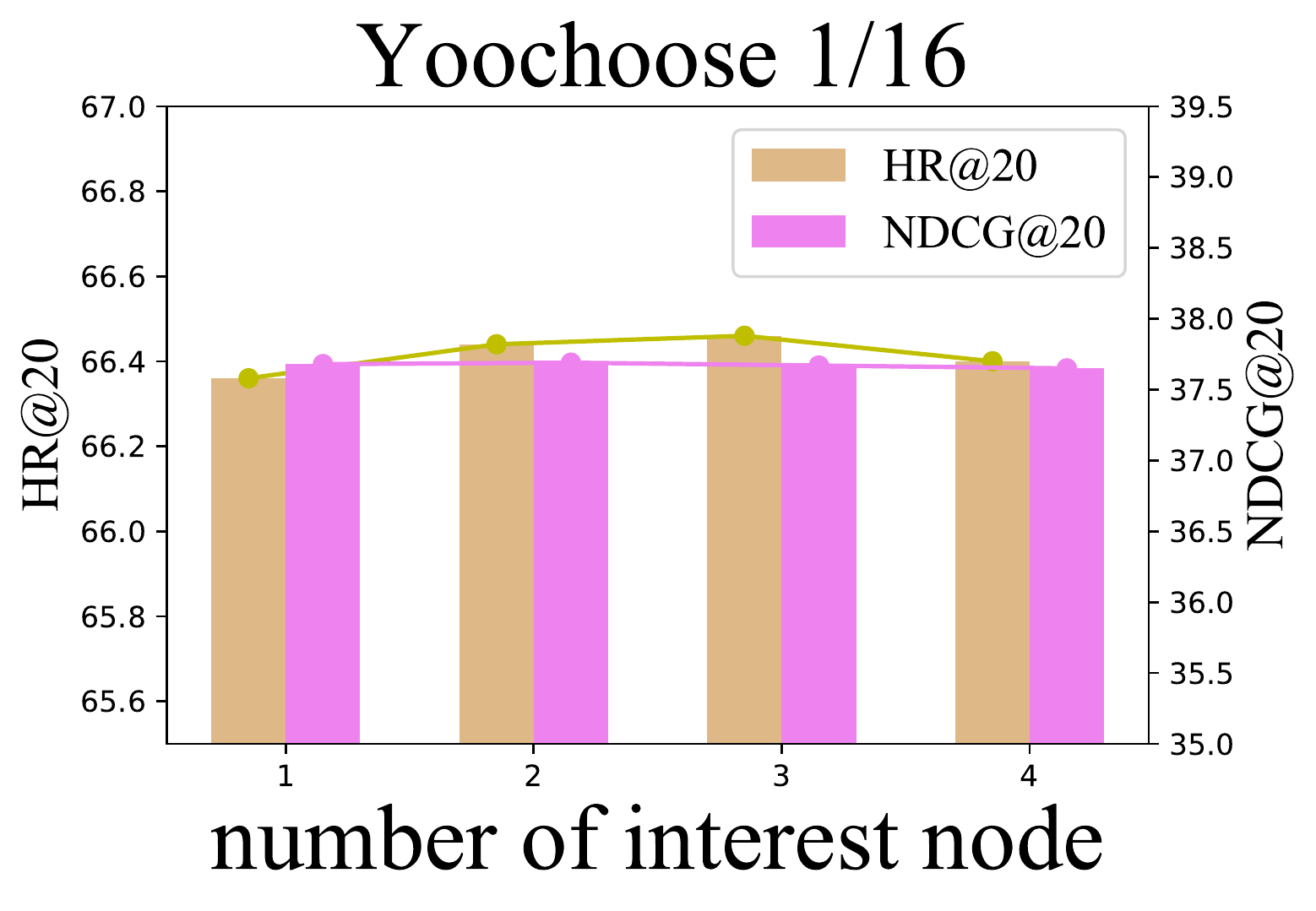}
    \end{subfigure}
    \begin{subfigure}{0.43\linewidth}
        \includegraphics[width=\textwidth]{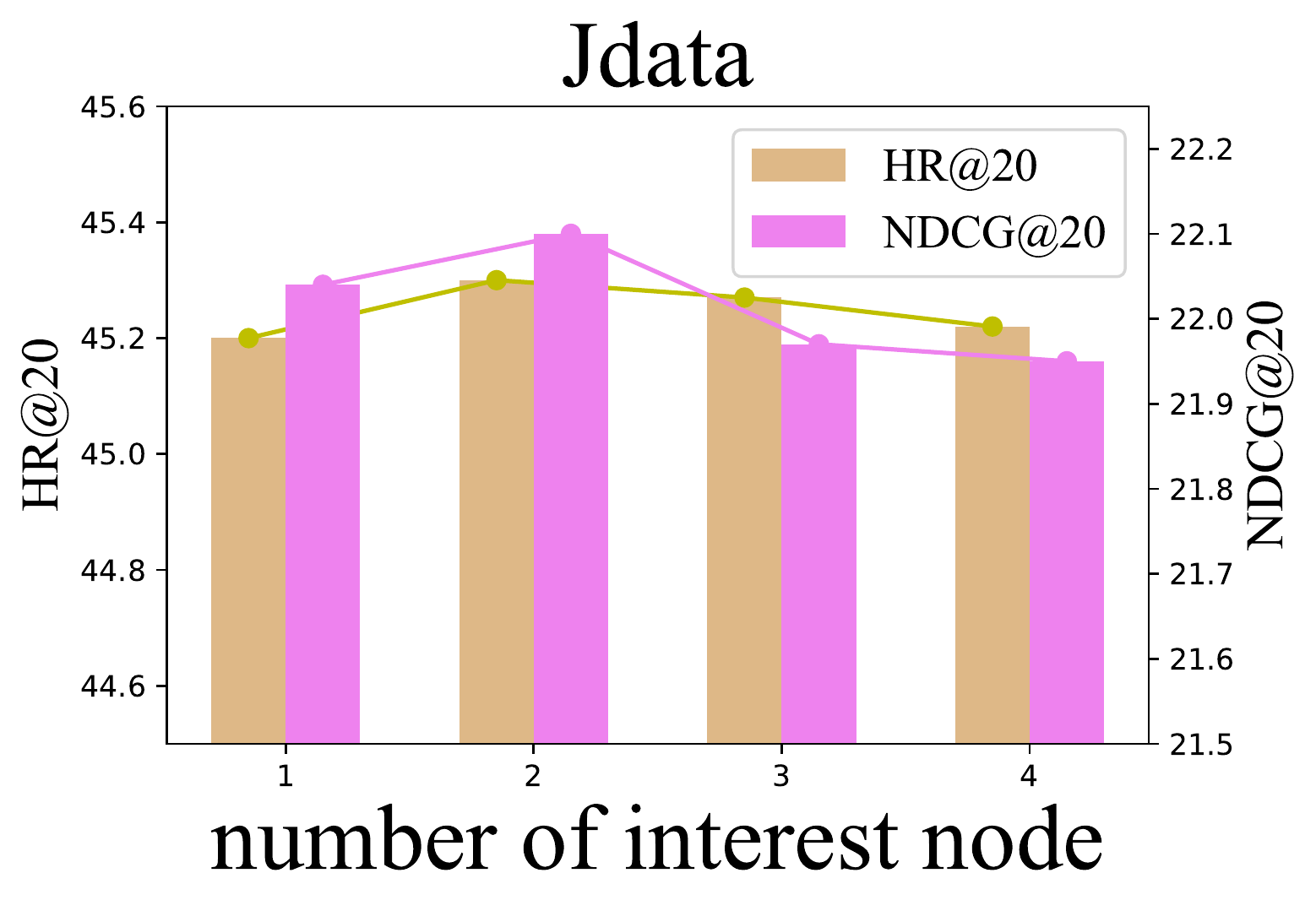}
    \end{subfigure}
    \caption{Performance comparison w.r.t. different time bucket width and max interest node number.}
    \label{fig:hyper-timebucket}
    \vspace{-0.4cm}
\end{figure}
As discussed in \autoref{sec:model}, we divide the time interval into buckets to utilize time signals.
So we conduct tests by ranging the time bucket width within $\{2,4,8,16,32\}$ to explore how does the bucketing setting affect the model's performance.
As shown in figure \ref{fig:hyper-timebucket}, we can see that the performance does not fluctuate dramatically as the time bucket width changes, which indicates that our model is not sensitive to the time bucket width.

\noindent \textbf{Impact of interest node num.}
% \subsubsection{Impact of Max Interest Node Num}
To investigate the impact of the interest node number, we range this parameter in $\{1,2,3,4\}$. According to the results in \autoref{fig:hyper-timebucket}, we can see that for both \emph{Yoochoose} and \emph{Jdata}, the model with 2 interest nodes achieves the best performance in most metrics. 
Compare to single interest node with mixed interest representation, our model with two interest nodes disentangles the latent interest representation for better next item prediction.
When the number becomes larger, performance will drop due to the redundancy of interest representation. 
% Besides, the performance 

%The comparation indicates that for small dataset, our model tends to capture simple interest distribution pattern which is divided by relatively long time interval. 
%For large amount of data, our model can learn more sophisticated interest distribution patterns, so more detailed time bucketing contributes to better  result.

%  on dataset \emph{Yoochoose 1/16}, setting time bucket width to 8 seconds brings the best result, while the best width is 2 seconds on \emph{Jdata}.

% \begin{figure}[t]
%     \centering
%     \includegraphics[width=0.45\textwidth]{img/case_study_2.pdf}
%     \caption{A case study of two users from \emph{Last.fm} data for music artist recommendation. Two users have different music preferences but similar session sequences, which are the basis to generate the recommendation. This figure presents the difference between the recommendation results (top 5 artists) generated by our model and the base model.}
%     \label{fig:case_study)}
% \end{figure}

% \subsection{Case Study (RQ4) }\label{sec:case_study}
% To illustrate the 

\vspace{-0.15cm}
\section{Conclusion}
\vspace{-0.1cm}
In this paper, we pay special attention to the disentangled  multi-interest representations of user and multiple temporal information for session based recommendation.
% emph{Temporal aware Multi-Interest Graph Neural Network} 
We construct a multi-interest graph and devise the TMI-GNN model, which utilizes the multi-interest graph to capture adjacent item transitions, distill multi-interest representations with the injection of multi-form temporal information.
In the experiments, our model outperforms other state-of-the-art session-based models, showing the effectiveness of our model.

%future work
% This work shows initial attempts towards disentangled multi-interest for SBR. 
% In future work, we will explore other side information, such as item side  knowledge and user reviews to further interpret disentangled interest representations better.

%%
%% The next two lines define the bibliography style to be used, and
%% the bibliography file.
%\clearpage
\bibliographystyle{ACM-Reference-Format}
\bibliography{ref}

%%
%% If your work has an appendix, this is the place to put it.
% \appendix

% \input{appendix.tex}

\end{document}